\documentclass[fleqn,usenatbib]{mnras}

\usepackage{newtxtext,newtxmath}

\usepackage[T1]{fontenc}
\usepackage{ae,aecompl}

\usepackage{graphicx}	
\usepackage{amsmath}	
\usepackage{cleveref}   
\usepackage{siunitx}
\usepackage{subcaption}

\crefformat{section}{\S#2#1#3} 
\crefformat{subsection}{\S#2#1#3}
\crefformat{subsubsection}{\S#2#1#3}

\title[On-Sky Testing of the iLocater Acquisition Camera]{Final Design and On-Sky Testing of the iLocater SX Acquisition Camera: Broadband Single-Mode Fiber Coupling}

\author[J. Crass et al.]{
J. Crass,$^{1}$\thanks{E-mail: j.crass@nd.edu}
A. Bechter,$^{1}$
B. Sands,$^{2}$
D. King,$^{3}$
R. Ketterer,$^{1}$
M. Engstrom,$^{2}$
\newauthor
R. Hamper,$^{2}$
D. Kopon,$^{4}$
J. Smous,$^{2}$
J. R. Crepp,$^{1}$
M. Montoya,$^{5}$
O. Durney,$^{5}$
\newauthor
D. Cavalieri,$^{2}$
R. Reynolds,$^{6}$
M. Vansickle,$^{1}$
E. Onuma,$^{7}$
J. Thomes,$^{7}$
S. Mullin,$^{2}$
\newauthor
C. Shelton,$^{8}$
K. Wallace,$^{8}$
E. Bechter,$^{1}$
A. Vaz,$^{5}$
J. Power,$^{6}$
G. Rahmer,$^{6}$
and S. Ertel$^{6,5}$
\\
$^{1}$Department of Physics, University of Notre Dame, 225 Nieuwland Science Hall, Notre Dame, IN 46556, U.S.A.\\
$^{2}$Engineering and Design Core Facility, University of Notre Dame, Notre Dame, IN 46556, U.S.A.\\
$^{3}$Institute of Astronomy, University of Cambridge, Madingley Rd, Cambridge, CB3 0HA, U.K. \\
$^{4}$Harvard Smithsonian Center for Astrophysics, 60 Garden St., Cambridge, MA 02138, U.S.A.\\
$^{5}$Department of Astronomy/Steward Observatory, University of Arizona, 933 N. Cherry Ave, Tucson, AZ 85721-0009, U.S.A.\\
$^{6}$Large Binocular Telescope Observatory, 933 N. Cherry Ave, Tucson, AZ 85721-0009, U.S.A.\\
$^{7}$NASA Goddard Space Flight Center, 8800 Greenbelt Rd, Greenbelt, MD 20771, U.S.A.\\
$^{8}$Jet Propulsion Laboratory, California Institute of Technology, 4800 Oak Grove Drive, Pasadena, CA 91125, U.S.A.\\
}

\date{Accepted 2020 October 22. Received 2020 October 21; in original form 2020 September 24}

\pubyear{2020}

\begin{document}
\label{firstpage}
\pagerange{\pageref{firstpage}--\pageref{lastpage}}
\maketitle

\begin{abstract}
Enabling efficient injection of light into single-mode fibers (SMFs) is a key requirement in realizing diffraction-limited astronomical spectroscopy on ground-based telescopes. SMF-fed spectrographs, facilitated by the use of adaptive optics (AO), offer distinct advantages over comparable seeing-limited designs, including higher spectral resolution within a compact and stable instrument volume, and a telescope independent spectrograph design. \emph{iLocater} is an extremely precise radial velocity (EPRV) spectrograph being built for the Large Binocular Telescope (LBT). We have designed and built the front-end fiber injection system, or acquisition camera, for the SX (left) primary mirror of the LBT. The instrument was installed in 2019 and underwent on-sky commissioning and performance assessment. In this paper, we present the instrument requirements, acquisition camera design, as well as results from first-light measurements. Broadband single-mode fiber coupling in excess of 35\% (absolute) in the near-infrared (\SIrange{0.97}{1.31}{\micro\meter}) was achieved across a range of target magnitudes, spectral types, and observing conditions. Successful demonstration of on-sky performance represents both a major milestone in the development of \emph{iLocater} and in making efficient ground-based SMF-fed astronomical instruments a reality.
\end{abstract}




\section{Introduction}

Precise Doppler radial velocity (RV) instruments continue to play an essential role in the study of exoplanets \citep{NationalAcademiesofSciencesEngineering2018}. Current RV instruments reach a precision limit of approximately \SI{1}{\metre\per\second}, where asymmetries and changes in absorption line profiles caused by host star variability dominate RV planetary signals \citep{Cegla2013,Oshagh2017,Cegla2018}. Applied theoretical work has shown that high spectral-resolution ($R>150,000$) and high signal-to-noise measurements are needed to mitigate the effects of stellar variability \citep{Davis2017}. However, achieving this capability with seeing-limited instruments illuminated by multi-mode fibers (MMFs) is challenging. Modal-noise from MMFs creates illumination instabilities. Additionally, obtaining high signal-to-noise measurements at high spectral-resolution requires large collecting areas, which in the seeing-limited regime, requires significant instrument volumes that are expensive to build and difficult to stabilize \citep{Strassmeier2008,Crepp2014,Crass2019}. It is therefore prudent to develop new techniques to overcome current precision limitations and enable efficient detection of sub-m/s RV amplitude exoplanets.

Advancements in adaptive optics (AO) systems and associated technologies mean that today, AO-equipped ground-based telescopes are able to deliver near diffraction-limited beams to instruments. Correction capabilities are now routine at near-infrared (NIR) wavelengths and continue to improve towards shorter, visible wavelengths \citep{Poyneer2014,Jovanovic2015,Males2018,Beuzit2019}. With routine AO performance comes the option to fundamentally re-assess instrument designs from their seeing-limited versions to those fed by a diffraction-limited input. The use of AO enables the efficient injection of starlight into significantly smaller optical fibers than current instruments. As the fiber diameter decreases, fibers reach the limit where they only support light propagation in a single spatial mode, i.e. single-mode fibers (SMFs). Over the past several years, the use of SMFs in ground-based astronomical instruments has continued to grow with several instruments now using them for illumination \citep{Blake2015,Crepp2016,GravityCollaboration2017,Feger2018,Mawet2019}.

A SMF operates at the diffraction limit, propagates light at wavelengths similar to the fiber glass core diameter, and outputs a spatially stable Gaussian beam profile. This stability mitigates illumination errors including modal-noise effects in MMFs, and enables the use of photonic technologies in next-generation instrument development \citep{BlandHawthorn2004,Morris2016,Anagnos2018,Jovanovic2020}. If used to illuminate a precision spectrograph, operating in the diffraction-limited regime removes the relationship between telescope diameter and spectrograph design experienced by seeing-limited instruments. This enables a high-spectral resolution instrument design to be achieved in a compact instrument volume for any telescope diameter. The decoupling of telescope aperture and spectrograph design allows effective pathways to develop SMF-fed instruments that are compatible with both current and upcoming large ground-based facilities. To enable this innovation however, efficient injection of light into these SMFs must be achieved on ground-based facilities \citep{Crepp2014,Jovanovic2014,Jovanovic2016,Jovanovic2017}.

iLocater is an extremely precise radial velocity (EPRV) instrument being developed for the dual \SI{8.4}{\metre} Large Binocular Telescope (LBT) \citep{Crepp2016}. The instrument front-end system, or acquisition camera, is fed using the LBT AO system. The acquisition camera comprises two independent modules, one for each telescope primary, with each module receiving the AO-corrected beam from one of the telescope primary mirrors. Light is efficiently coupled into a $\sim$\SI{6}{\micro\metre} diameter SMF with imaging of the target being provided simultaneously.

In this paper, we present the final design and first light on-sky tests of the iLocater SX acquisition camera which couples light from the left (SX) primary mirror of the LBT. The work presented builds heavily upon the research and development undertaken to install a prototype system at the LBT in 2016 and characterize SMF coupling performance \citep{Bechter2015,Bechter2020}. We discuss the requirements needed for efficient SMF coupling (\cref{sec:efficientcoupling}) and how these are met through the acquisition camera optical (\cref{sec:optdesign}) and mechanical (\cref{sec:mechdesign}) design. Key instrument control features are also described (\cref{sec:control}). The SX acquisition camera module was installed at the LBT in June 2019 with on-sky commissioning taking place during July 2019 and November 2019. We present the laboratory performance of the system prior to shipment to the telescope (\cref{sec:labperf}) and its subsequent on-sky performance (\cref{sec:skyperf}). Finally, we conclude with a discussion of performance and prospective hardware upgrades (\cref{sec:conclusions}).
\section{Requirements to achieve efficient single-mode fiber coupling}
\label{sec:efficientcoupling}

Achieving efficient injection of light into SMFs requires careful control and matching of the incident optical beam onto the fiber tip. Efficient coupling is achieved when the incident beam has a flat wavefront and spatial profile which closely matches the fundamental propagation mode of the SMF itself. While strictly a Bessel function, in certain conditions, for example, in the regime close to the single-mode cutoff wavelength, this propagation mode can be approximated as a Gaussian profile \citep{Neumann2018}. If the fiber is considered as a spatial filter approximated by this Gaussian profile, for a diffraction-limited incident beam, coupling efficiency can to first order be determined by computing the overlap integral at the focal plane between the fiber spatial profile and the electric field distribution of the incident beam \citep{Shaklan1988, Ruilier1998}. Using this methodology, the spatial size of the fiber and incident beam must be well matched to achieve efficient coupling. In previous work, spatial matching is often considered in the context of optimizing the $f$-number of the telescope and directly coupling into fibers at the focal plane. However, in practical applications, the incident beam needs to be re-imaged using an optical system to achieve optimal coupling as a typical telescope design is not optimized for this application.

\subsection{Optical Design Optimization}

For a diffraction-limited input represented by an Airy disk (i.e. a perfect circle with no obscuration), the optimal sizing is found when the $1/e^2$ diameter of the incident beam is approximately matched to the mode-field diameter (MFD) of the fiber, defined as the $1/e^2$ width of its fundamental mode profile \citep{Ruilier2001}. For full optimization or when illuminating using more complex profiles, for example those which may no longer be considered diffraction-limited or which have a non-circular aperture, it is easier to assess the expected coupling performance in a pupil plane using the overlap of the complex electric field of the incident beam and fiber Gaussian profile. In ground-based astronomical applications, the telescope primary mirror typically defines the diffraction profile and spatial scale at the telescope focal plane. This is modified by any items occulting the beam, for example a secondary mirror and its support structures. Therefore, optimal overlap for coupling must be assessed on a specific telescope design accounting for its design features. A secondary mirror obscuration of less than 20\% of the telescope primary typically has a minimal effect on coupling performance \citep{Ruilier1998}.

While it may be straight-forward to consider the case of monochromatic illumination of a SMF, broadband illumination, as used in astronomical applications, is more challenging. Careful consideration of wavelength-dependent effects is necessary, such as growth in diameter of the incident beam with wavelength due to diffraction, refractive effects of the atmosphere and optical components, the wavelength-dependent performance of AO systems, and the change in diameter of the fundamental mode of the optical fiber with wavelength. Additionally, assessing the short-term and long-term stability of these effects is critical in achieving optimal SMF coupling.

\begin{figure}
    \centering
    \begin{minipage}{\columnwidth}
        \centering
        \includegraphics[clip, trim=2.0cm 1cm 2.0cm 1.4cm, width=1.0\columnwidth]{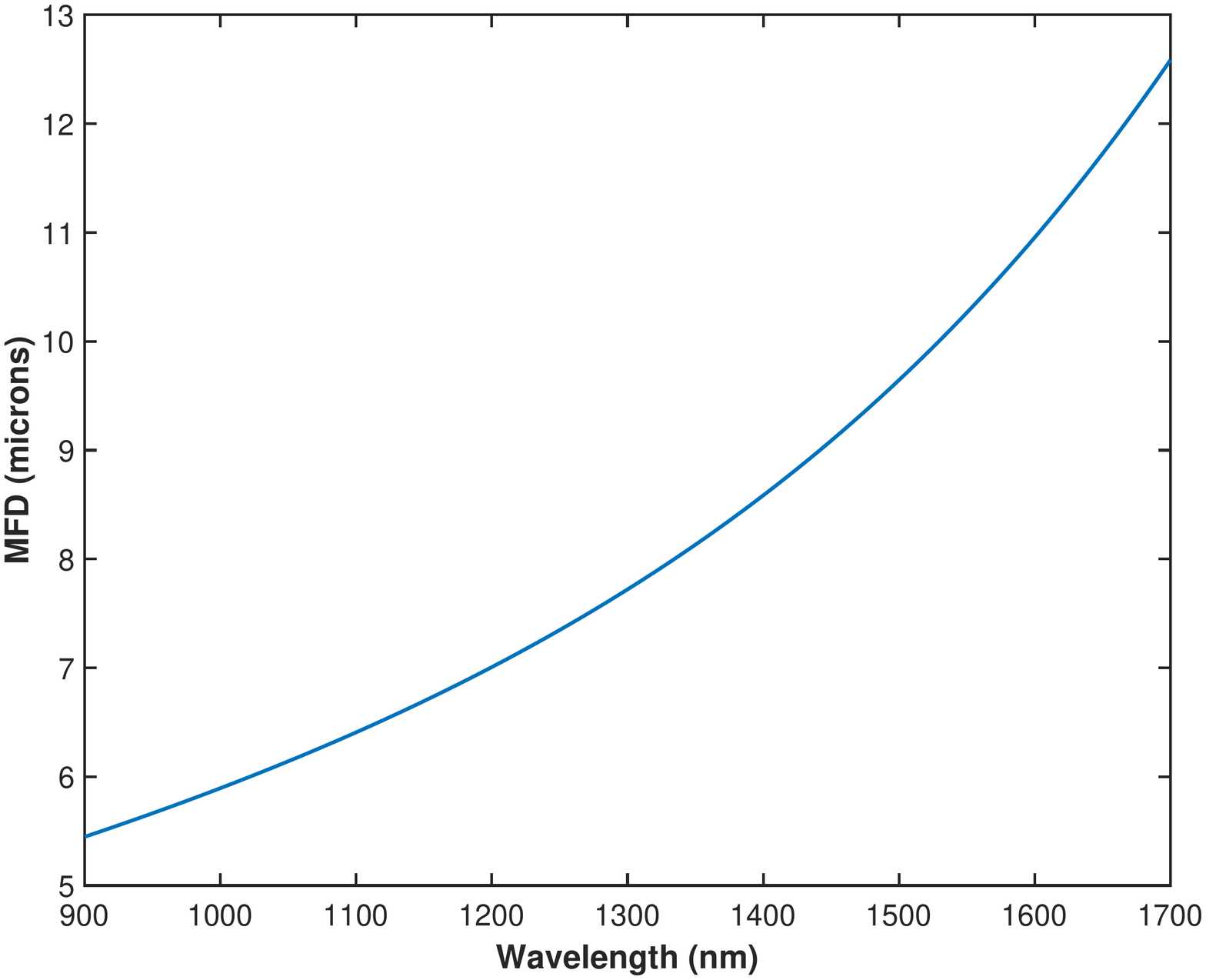}
        \end{minipage}\hfill
    \begin{minipage}{\columnwidth}
        \centering
        \includegraphics[clip, trim=2.0cm 1cm 2.0cm 1.0cm, width=1.0\columnwidth]{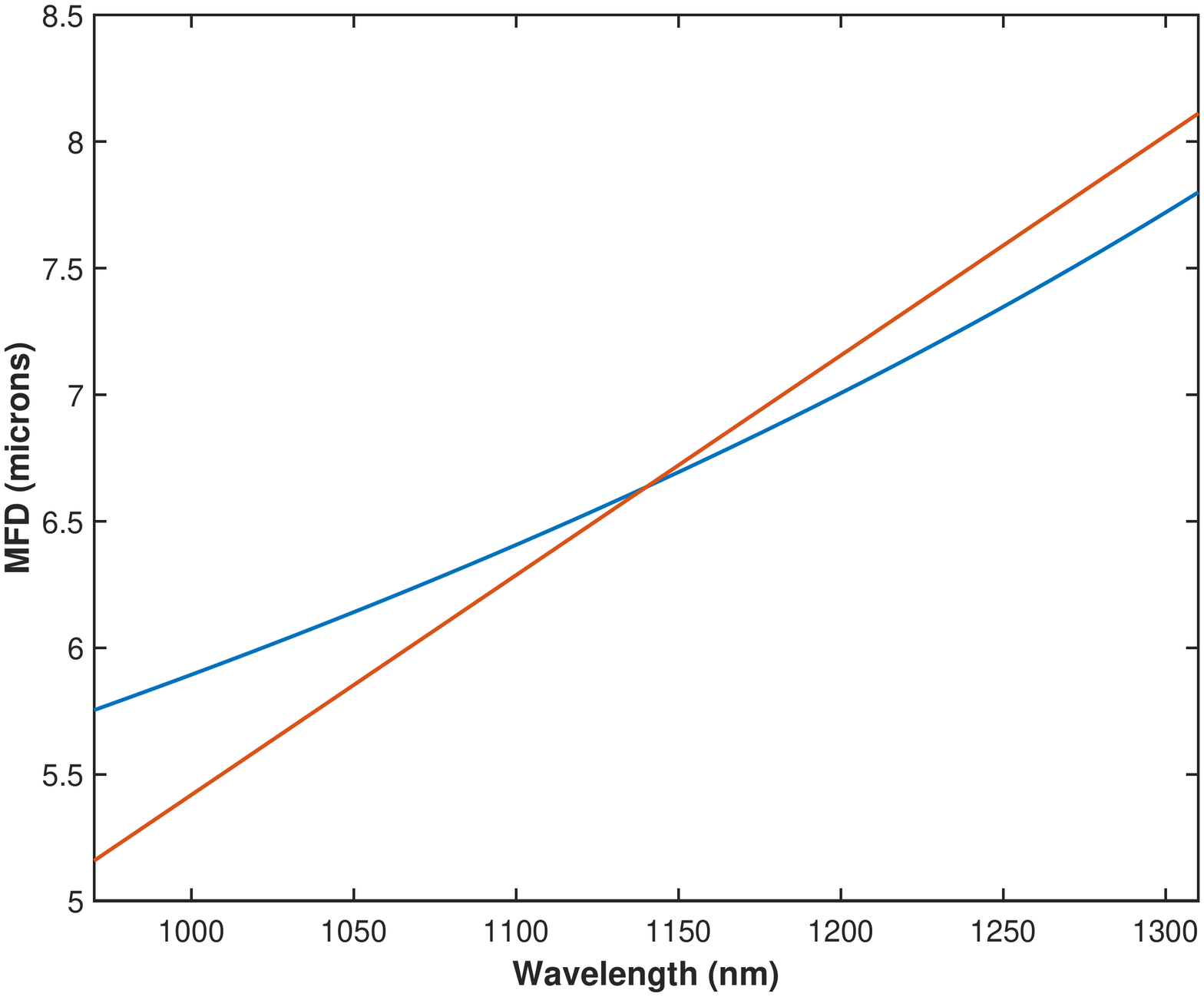}
    \end{minipage}
   \caption{Calculated dependence of the MFD with wavelength (blue) for the SMF used in iLocater (SM980). The lower figure shows a zoomed in region of wavelength coverage which matches the iLocater bandpass. A linear relationship representing diffraction effects from the LBT telescope input is plotted for reference (orange) which is centered on the midpoint of the instrument wavelength range.}
    \label{fig:MFD}
\end{figure}

For SMFs, the dependence of MFD on wavelength is governed by the radial extent of the fundamental mode of the electromagnetic field. Several factors impact the wavelength dependence including the fiber physical structure (e.g. a step index fiber comprising a core and cladding of different refractive indices, or a graded fiber where the refractive index varies smoothly in the radial direction), dispersion, and any imperfections such as bends. This leads to numerous prescriptions depending on the precise fiber design, and for a situation requiring precise understanding of the wavelength dependence, it is often simplest to empirically determine these properties. In certain circumstances, approximations can be made to determine a simple relationship between MFD and wavelength. For example, when a step index fiber is illuminated with wavelengths within a limited regime from the single-mode cutoff wavelength (approximately within a factor of two), the MFD will grow proportionally with wavelength \citep{Neumann2018}. Dependence of the MFD for the SMF used in iLocater is shown in Fig.~\ref{fig:MFD}. 

While the diameter of a diffraction-limited telescope input beam grows linearly with wavelength, the MFD of a fiber grows more rapidly. If however, a narrow wavelength range close to the SMF short wavelength cutoff is used and the telescope input size is optimized (to first order at the center of the wavelength band), the divergence between these two functions is small, and therefore efficient coupling can be achieved across the band. Given the practical effects of an imperfect incident telescope beam, the mismatch between the MFD with a diffraction-limited incident point spread function (PSF) typically has much less of an impact on performance than other dynamic effects which often severely limit coupling efficiency.

\subsection{Optimizing for Dynamic Effects}

While optimization of the theoretical incident beam of the telescope is critical for system design, to achieve efficient coupling in a practical situation, it is important to consider dynamic changes that can impact performance. In ground-based astronomy applications, we are attempting to align a broadband input beam which has passed through the atmosphere, inducing wavefront distortions and chromatic effects, onto a SMF. Any deviations from the optimal incident beam will reduce fiber injection efficiency, in some cases significantly \citep{Ruilier1998,CoudeDuForesto2000,Woillez2004,Bechter2020}. Therefore, it is critical to mitigate any dynamic effects in the instrument system upstream of the fiber input and to maintain a stable beam.

The distorting effects of atmospheric turbulence can be characterized in terms of the Fried Parameter, $r_0$. This is a measure of the atmospheric coherence and can be considered as the diameter of an air cell at a telescope aperture which is unaffected by turbulent effects. A larger telescope has many more turbulent cells impacting an incident beam and in the context of AO performance, each cell requires a corrective element to achieve effective correction. As $r_0$ scales as $\lambda^{6/5}$, at shorter wavelengths each cell is smaller and therefore for the same telescope aperture more elements are required for correction. Combined with the faster correction timescales needed to account for smaller cells moving across the telescope, this leads to more complex AO systems for larger telescopes and a performance degradation at shorter wavelengths. It is important therefore to ensure that any AO system feeding a SMF is able to provide adequate correction performance across the entire operating wavelength range. The Strehl ratio can serve as a first order estimator for expected fiber coupling performance when fed by a specific AO system, however, this fails to fully capture the sensitivity to specific residual aberrations and therefore, for a true picture, a more rigorous analysis is required \citep{Ruilier1998}.

As AO systems often sense beam distortions well upstream of the final science instrument, it is important to ensure the system is capable of correcting for non-common path aberrations (NCPA) induced on the downstream beam. Depending on the telescope and instrument design, these aberrations can change with telescope elevation or environmental conditions and therefore any system should incorporate an effective method to determine, characterize and correct for NCPA. 

Refraction chromatically disperses starlight as it passes through the atmosphere. To achieve efficient SMF coupling, all wavelengths must be co-aligned at the fiber input and therefore a precise method to correct for atmospheric refraction is required. Typically this is achieved using a prism-based atmospheric dispersion corrector (ADC). For a diffraction-limited system, even residuals of corrected dispersion must be minimized including secondary effects arising from wavelength dependent refractive index in any prism glasses used. Any residual dispersive effects from optical elements, e.g. angled entrance windows, also need to be considered in this correction process to ensure optimal performance.

The complex nature of telescope systems and their exposure to the elements make them potential sources of vibrations on instrument support structures. Vibrations can induce tip/tilt errors that are not sensed or corrected by AO systems and without additional correction can lead to dynamic mis-alignment when attempting SMF coupling. Such tip/tilt residuals must be handled in real-time to maintain beam alignment and efficient SMF coupling.
\section{Optical Design}
\label{sec:optdesign}

\begin{figure}
	\includegraphics[width=\columnwidth]{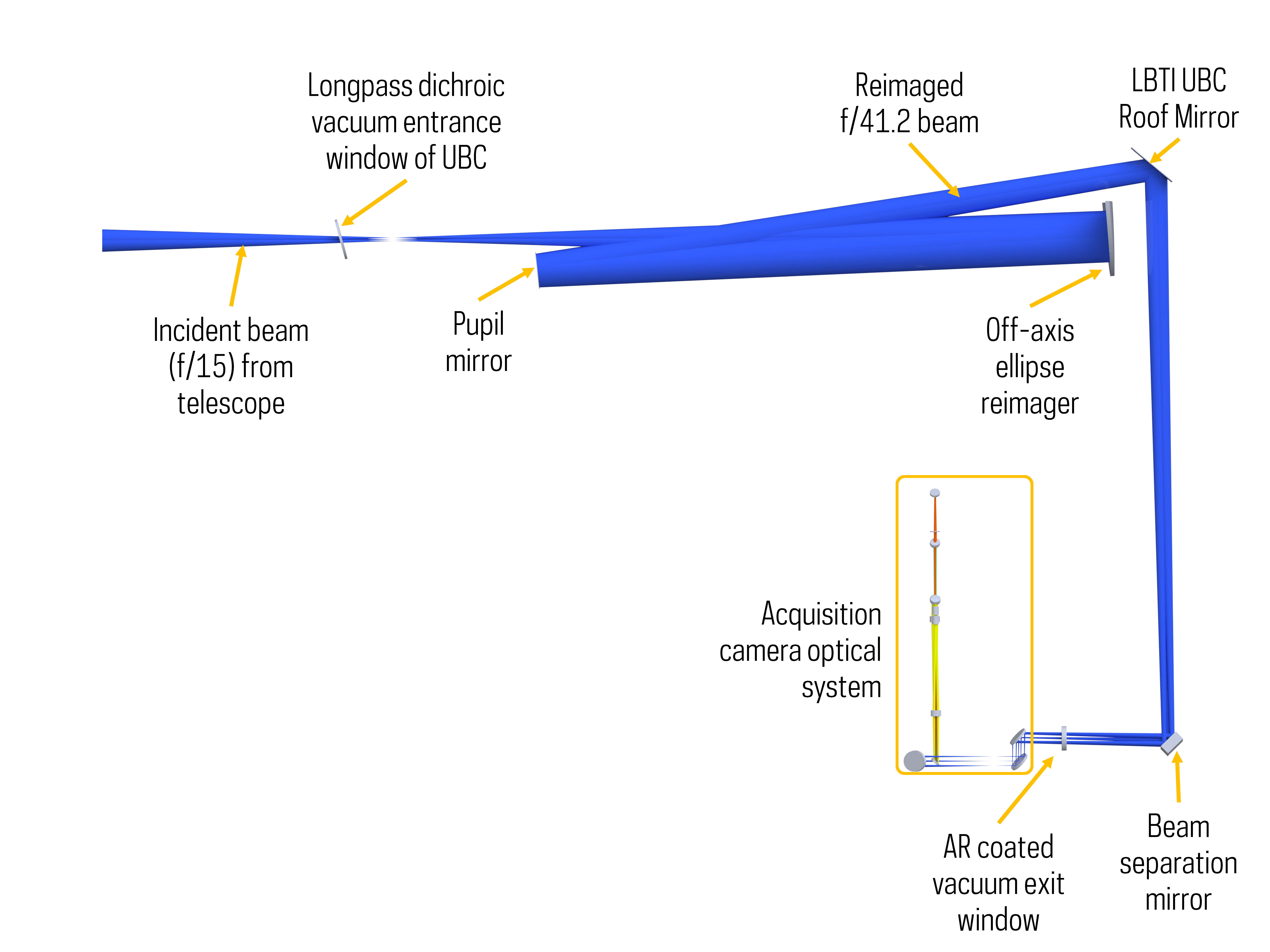}
    \caption{Optical layout of the upstream optical components feeding the iLocater SX acquisition camera. The f/15 beam from the telescope is split by wavelength using a dichroic entrance window at the entrance of LBTI. Visible light is reflected to the LBTI wavefront sensor while science light enters the Universal Beam Combiner (UBC) and is reimaged to f/41.2. The beam is then steered using internal UBC optics before being separated from the other telescope beam (DX) and exiting the UBC system and entering the SX acquisition camera module.}
    \label{fig:optical_design_lbti}
\end{figure}

The iLocater acquisition camera has been designed to meet the requirements outlined for efficient SMF coupling. The system is fed using incident beams from the Large Binocular Telescope Interferometer (LBTI) (Fig.~\ref{fig:optical_design_lbti}) which is installed on the center bent Gregorian focal station of the LBT \citep{Hinz2004,Cheng2004,Hinz2016}. LBTI was developed to receive the beams from both LBT primary mirrors for a range of high angular resolution and high contrast observing modes, and precise beam overlapping for interferometry \citep{Stone2018, Spalding2018, Ertel2018, Ertel2020}. It uses the LBT's adaptive secondary mirrors and LBTI's pyramid wavefront sensors for closed-loop AO correction. The wavefront sensors have recently been upgraded to use OCAM2K detectors which improve noise performance and correction speeds, and increase both the limiting magnitude and AO correction, a critical requirement for efficient SMF injection \citep{Pinna2016, Pinna2019}. For iLocater, we utilize LBTI's AO correction and beam steering capabilities with its Universal Beam Combiner (UBC) to deliver the telescope beams to the SX (left) and DX (right) acquisition camera modules.

\subsection{Optical Design Overview}

The optical design of the iLocater acquisition camera, is driven at the highest level by the exoplanet detection and characterization science requirements of the instrument spectrograph. These define the required instrument bandpass which in turn drives the SMF characteristics required for operation and defines the optical prescription needed for optimal fiber coupling.

iLocater will operate in the Y- and J-bands which provide relatively clean atmospheric windows with limited telluric contamination. The bands are well optimized for studies of late-type stars, one of iLocater's key science goals, which peak in emission in the NIR. The final instrument spectrograph bandpass (\SIrange{0.97}{1.31}{\micro\meter}) was selected after optimizing for both spectrograph detector coverage and photon noise performance at the spectrograph focal plane. This bandpass was derived from simulations assessing the performance of the coupling efficiency within the fiber injection system accounting for AO system performance for differing spectral types and V-band magnitudes \citep{Bechter2018}.

\begin{figure}
	\includegraphics[width=\columnwidth]{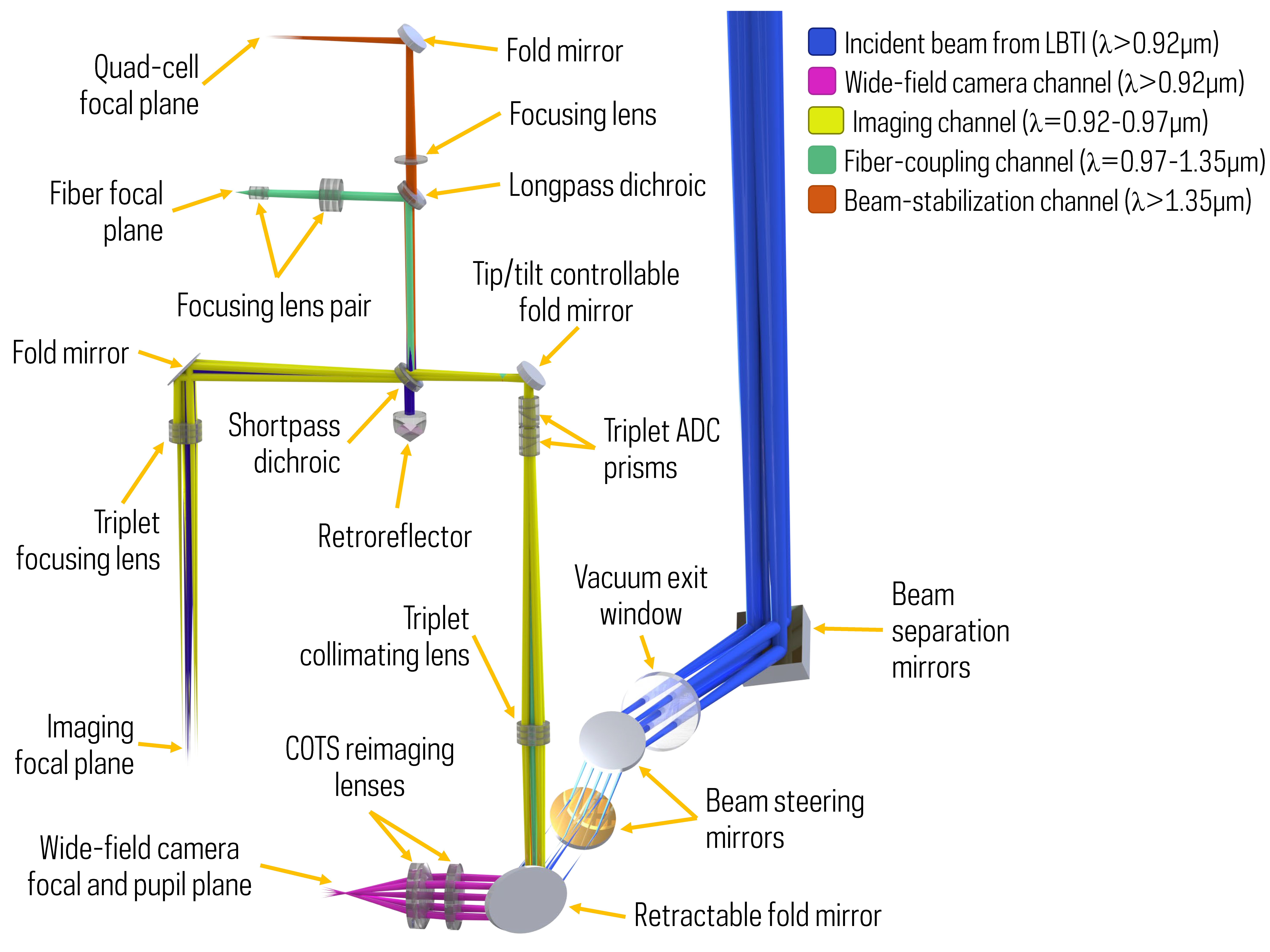}
    \caption{Optical design of the SX acquisition camera. The incident f/41.2 beam from LBTI is aligned to the optical axis of the iLocater system using a pair of steerable fold mirrors. Several channels are present within the acquisition camera, each serving a different purpose. A wide-field camera channel allows coarse acquisition to take place during alignment and by translating the detector along the optical axis, also provides pupil plane imaging. A mirror diverts the beam from the wide-field camera to illuminate the three other instrument channels which are generated using a long and shortpass dichroic in combination. These simultaneously illuminate an imaging detector, the fiber focal plane, and a quad-cell. An optically identical design is used for the DX system with some repositioning of channels and components.}
    \label{fig:optical_design}
\end{figure}

The internal optical design of the SX acquisition camera is shown in Fig.~\ref{fig:optical_design}. The incident beam from LBTI is divided into four channels: three simultaneously illuminated channels that provide target imaging, fiber-coupling, and beam-stabilization, and an on-demand wide-field camera (WFC) channel that provides coarse target acquisition and pupil imaging.

\subsection{Optical Fibers \& Fiber System}
\label{sec:fibers}

The fiber selected for use within iLocater is an SM980 variant from Fibercore. This model was chosen given the requirements for wavelength range, transmission, single mode operation cut-off, and performance at cryogenic temperatures within the spectrograph. A specific fiber batch was selected to ensure a single-mode cutoff below \SI{970}{\nano\meter}. This fiber batch has a MFD of \SI{5.8}{\micro\meter} at \SI{980}{\nano\meter} with a measured numerical aperture of 0.11. The fiber diameter including cladding is \SI{125}{\micro\meter}.

While it would be optimal for total throughput to construct the fiber system using a single continuous piece of fiber running end-to-end from the acquisition camera to the spectrograph, there were several concerns regarding this approach. Firstly, as the fiber must run into the instrument cryostat, any end-to-end system would have to include a vacuum feed-through assembly, leading to challenges for installing the fiber assembly. Additionally, due to the telescope environment, there were concerns that a single-assembly would increase the risk should a fiber break occur on the telescope structure. 
To mitigate these concerns, a fiber system comprising several sections was adopted. Each section was joined end-to-end using AVIM type connectors which are precision machined. When assembled correctly, these connectors offer excellent concentricity of a fiber core within the connector, repeatability and stability. This allows minimal losses at each connector interface with $>95\%$ throughput being routinely achieved.

At the acquisition camera fiber focal plane, a close-packed bundle of seven fibers is installed. The fiber bundle creates a ring of six outer fibers and a central fiber with the cores of neighboring fibers being separated by \SI{125}{\micro\meter}. The central fiber is connected to a \SI{40}{\meter} patch fiber to feed the instrument spectrograph while the six off-axis fibers can be reverse-illuminated for precise fiber positioning (\cref{sec:back-illumination}).
   
\subsection{Upstream optical train}

LBTI is fed with an f/15 beam from each telescope primary mirror (Fig.~ \ref{fig:optical_design_lbti}). The incident beam from each telescope is divided into two channels using a dichroic coated vacuum window at the entrance of the UBC. Visible wavelengths are reflected to the instrument wavefront sensors while longer wavelengths are transmitted inside the UBC. A ferrofluidic interchange mechanism allows different vacuum windows to be precisely positioned for specific instrument configurations. iLocater uses a custom fused silica vacuum entrance window that has been optimized to reflect wavelengths shorter than \SI{920}{\nano\metre} to the wavefront sensor with longer wavelengths up to \SI{2}{\micro\metre} being transmitted with $>96\%$ efficiency across the science bandpass. The use of fused silica for the entrance window introduces a slight beam deviation from the nominal LBTI optical design; however, this is compensated using the roof mirror beam steering capability within the UBC.

The beam from each telescope is reimaged to f/41.2 using an off-axis ellipse within the UBC before both independent telescope beams are brought to a common focal plane in the center of the UBC. When feeding light to iLocater, the telescope beams are separated using an angled pair of plane mirrors which are inserted within a newly installed instrument port selector. The plane mirrors direct each telescope beam through an anti-reflection (AR) coated vacuum exit window ($T>99\%$) to two independent fiber injection modules, one for each side of the telescope. The incident beam is then aligned to each module's optical axis using a pair of tip/tilt fold mirrors.

\subsection{Common optics}

The f/41.2 beam from LBTI passes through a set of common optics which feed the imaging, fiber-coupling, and beam-stabilization channels. Custom optics have been used to ensure diffraction-limited performance across all channels with overall performance being optimized for the fiber focal plane. End-to-end transmission is maximized using commercial off-the-shelf (COTS) protected gold coated mirrors and optimized AR coatings for the bandpass on all air-to-glass optical surfaces. A future upgrade includes replacing the COTS mirrors with multi-layer coated equivalents which are being delivered as part of the spectrograph optics fabrication. The upgraded mirrors are optimized for performance in the spectrograph bandpass and will increase the system throughput to the fiber focal plane by 4\%.

The incident beam is collimated using an f = \SI{257}{\milli\meter} custom triplet lens comprising Ohara glasses S-FPL51, S-BSL7 and SBAH28. The collimated beam passes through a pair of ADC prisms before being folded using a fast tip/tilt corrector mirror driven using the input signal from a quad-cell sensor in the beam-stabilization channel. The beam is split using a pair of custom fused-silica dichroics; The first is a shortpass dichroic (T = \SIrange{0.92}{0.95}{\micro\metre}, R = \SIrange{0.97}{1.76}{\micro\metre}) with the transmitted beam passing to the imaging channel while the reflected beam is split again using a longpass dichroic (T = \SIrange{1.34}{1.76}{\micro\metre}, R = \SIrange{0.97}{1.31}{\micro\metre}) which generates the reflected fiber channel and transmitted beam-stabilization channel.

\subsection{Atmospheric Dispersion Correction}

To correct for the effects of atmospheric dispersion and achieve broadband diffraction-limited imaging over the zenith distance \SIrange{0}{60}{\degree}, iLocater uses a pair of identical triplet prisms. The prisms have tilted surfaces to correct both primary and secondary color while minimizing beam deviation. The glasses used are Ohara S-PHM53, S-TIM8, and Schott N-KZFS4, all of which are readily available and have well-matched refractive indices and varying dispersions. N-KZFS4 is the anomalous dispersion glass giving leverage over secondary color to correct for the mismatch between the wavelength dependence of atmospheric dispersion and the glass refractive index. The overall design is similar to that deployed on the Magellan AO system \citep{Kopon2013}, with specific design parameters optimized for iLocater (LBT environment, glass types, ADC diameter, central thicknesses, and wedge angles).

For mechanical reasons, the ADC is situated a short distance before an intermediary pupil plane, however, it shows excellent performance over the science band. The front surface of the triplet prism is perpendicular to the incoming beam, and the rear surface is tilted so that the beam deviation is near zero at the fiber focal plane over the complete zenith distance range. The final design was reoptimized accounting for melt data of the specific glasses used in fabrication to ensure there is no motion in the image plane over all zenith and field positions. Imaging performance is diffraction-limited up to \SI{50}{\micro\meter} displacement in the fiber focal plane. The effect of varying atmospheric parameters (temperature $\pm$\SI{20}{\celsius}, pressure $\pm$\SI{20}{\milli\bar} and humidity \SIrange{5}{50}{\percent}) in the Zemax optical model of the system had no significant effect on ADC performance.

\subsection{Optical Channels}

Each optical channel within the acquisition camera is designed to optimize performance for its specific application. The imaging channel, generated from the transmitted beam of the shortpass dichroic, is folded and then brought to a focal plane using an f = \SI{257}{\milli\meter} triplet lens identical to the common optics collimator. This generates a one-to-one reimager of the incident f/41.2 beam from LBTI and delivers a $8\times$\ang{;;8} field of view with a design plate scale of \SI{0.607}{\arcsec\per\mm}, which is imaged onto a CMOS detector from ANDOR (Zyla 4.2 Plus). The optical system provides {\color{red}6.1} pixels across the FWHM of the diffraction-limited PSF.

The fiber-coupling channel is refocused using a pair of lenses. The first lens is another f = \SI{257}{\milli\meter} triplet identical to the common optics collimator and the second is an f = \SI{20}{\milli\meter} triplet. The second triplet lens comprises S-LAL14/BaF2/S-BSL7. BaF2 is not a particularly robust material, however, replacing this element with CaF2 which is more stable did not deliver the imaging performance required for the fiber feed. Sandwiching the BaF2 between the two glasses helps protect the optical surfaces from the telescope environment and the complete triplet is edge sealed for further protection. The two lenses image the beam onto the fiber surface at f/3.7 with a $1/e^2$ diameter of \SI{5.8}{\micro\meter} matching the MFD of the fiber at a wavelength of \SI{0.97}{\micro\meter}. The $f$-number can be adjusted to f/4.2 by changing the lens separation. It is important that the chromatic focal shift is minimized when feeding the beam into the single-mode fiber to allow efficient broadband coupling, hence it was necessary to use triplet rather than doublet lenses.

The beam-stabilization channel is generated using the transmitted beam from the longpass dichroic. The beam is focused using a COTS f = \SI{200}{\milli\meter} singlet lens (Thorlabs LA1708-C) and is then folded onto the instrument quad-cell.

All of the optical channels are fed using the common optics tip/tilt mirror which applies the beam-stabilization correction. To ensure a consistent correction across all channels, particular attention was given during instrument integration to the precise positioning of the hardware at each channel's focal plane. As the command signal for tip/tilt correction is generated from the quad-cell in the beam-stabilization channel, if this sensor is offset from the focal plane, the other instrument channels will observe a beam wander as the tip/tilt correction is applied. In the case of the fiber-coupling channel, any observed beam wander will reduce coupling efficiency. Conversely, if the imaging detector and fiber are not located precisely at their focal planes, a similar effect will be noted if the quad-cell was positioned correctly. Careful attention was paid to this effect during the instrument integration to ensure accurate placement of these elements at their focal planes.

\subsection{Alignment and operation channels}

Several secondary optical channels are included within the optical design to aid with acquisition of the incident optical beam, coupling light into the instrument fiber (\cref{sec:fiber_positioning}), and enabling injection of calibration light to the instrument spectrograph.

\subsubsection{Wide-field camera}

The WFC channel provides coarse target acquisition and is illuminated by removing the fold mirror which feeds the common optics. The beam is focused using a pair of \SI{50.4}{\mm} diameter COTS doublets (f = \SI{250}{\milli\meter}, Thorlabs AC508-250-b and f = \SI{80}{\milli\meter}, Thorlabs AC508-80-b) which generates a $14\times$\ang{;;14} field of view on a GigE camera from Basler (acA2500-20gm). The WFC channel does not include any atmospheric dispersion correction, however, this does not impact the ability to provide coarse acquisition. A pupil plane is generated \SI{19}{\milli\meter} beyond the focal plane which can be accessed by scanning the Basler camera along the channel optical axis. 

\subsubsection{Fiber reverse-illumination}
\label{sec:back-illumination}

To couple incident starlight to the instrument fiber (\cref{sec:fiber_coupling}), the position of the fiber must be known relative to the telescope beam. This is accomplished by providing a simultaneous optical reference of the fiber location and telescope beam within the imaging channel.

The optical reference for alignment is provided by the off-axis fibers in the fiber bundle (\cref{sec:fibers}) which are back-illuminated from the focal plane with quasi-monochromatic light ($\lambda\sim$\SI{1.05}{\micro\meter}) from a superluminescent light emitting diode (sLED). Light is transmitted through the fiber-coupling channel triplets, reflected from the longpass dichroic, and $0.5\%$ is transmitted through the shortpass dichroic. A retroreflector is used to direct the beam back towards the shortpass dichroic where it is reflected into the imaging channel and imaged onto the ANDOR camera. iLocater uses three off-axis fibers for the reference illumination process. The reference fibers surround the science fiber and are optimized to produce approximately even illumination.

\subsubsection{Calibration Channel}

All illumination traces within the instrument spectrograph must be precisely calibrated during science observations. A calibration beam can be injected into the acquisition camera optical path just before the ADC prisms. The beam is generated from a calibration fiber coupled to an external source which is collimated using a COTS f = \SI{25}{\milli\meter} achromat (Thorlabs AC127-025-B) and then directed into the optical train using a pair of retractable fold mirrors. The beam then follows the normal path to the SMF focal plane.

\subsection{Tolerancing}

Tolerancing of the optical design has focused on the elements feeding the fiber focal plane as the combination of the f = \SI{257}{\milli\meter} and f = \SI{20}{\milli\meter} triplet lenses, which re-image the beam onto the fiber surface, are the most sensitive in the design. A standalone Zemax model was built to perform the tolerance analysis. Parameters analyzed include optical surface curvatures, surface irregularities, thicknesses between optical surfaces, decentering of surfaces and elements{\color{red},} and tilt of elements. The distance from the final lens to the fiber focal plane represented the final compensator in the analysis. 

Tolerance analysis was completed by setting minimum and maximum deviations for each of the above operands from the nominal value and by setting relevant compensators. A merit function was generated for the design and used in combination with the operands. Using a sensitivity analysis approach, each operand was adjusted to the minimum deviation value, the model was optimized using the defined compensators, and the new merit function calculated. This process was repeated for the same operand set to the maximum deviation value. Each operand in turn was dealt with in this way with the performance degradation being inspected and the relevant tolerance adjusted as necessary. This procedure quickly enables adjustment of the operand tolerances to keep the optical performance within the allowed margin. Operand tolerances were set iteratively until individual contributions to the RMS spot radius were each below \SI{1}{\micro\meter}. 

\section{Mechanical Design}
\label{sec:mechdesign}

\begin{figure}
    \centering
    \begin{subfigure}{\columnwidth}
        \centering
        \includegraphics[width=1.00\textwidth]{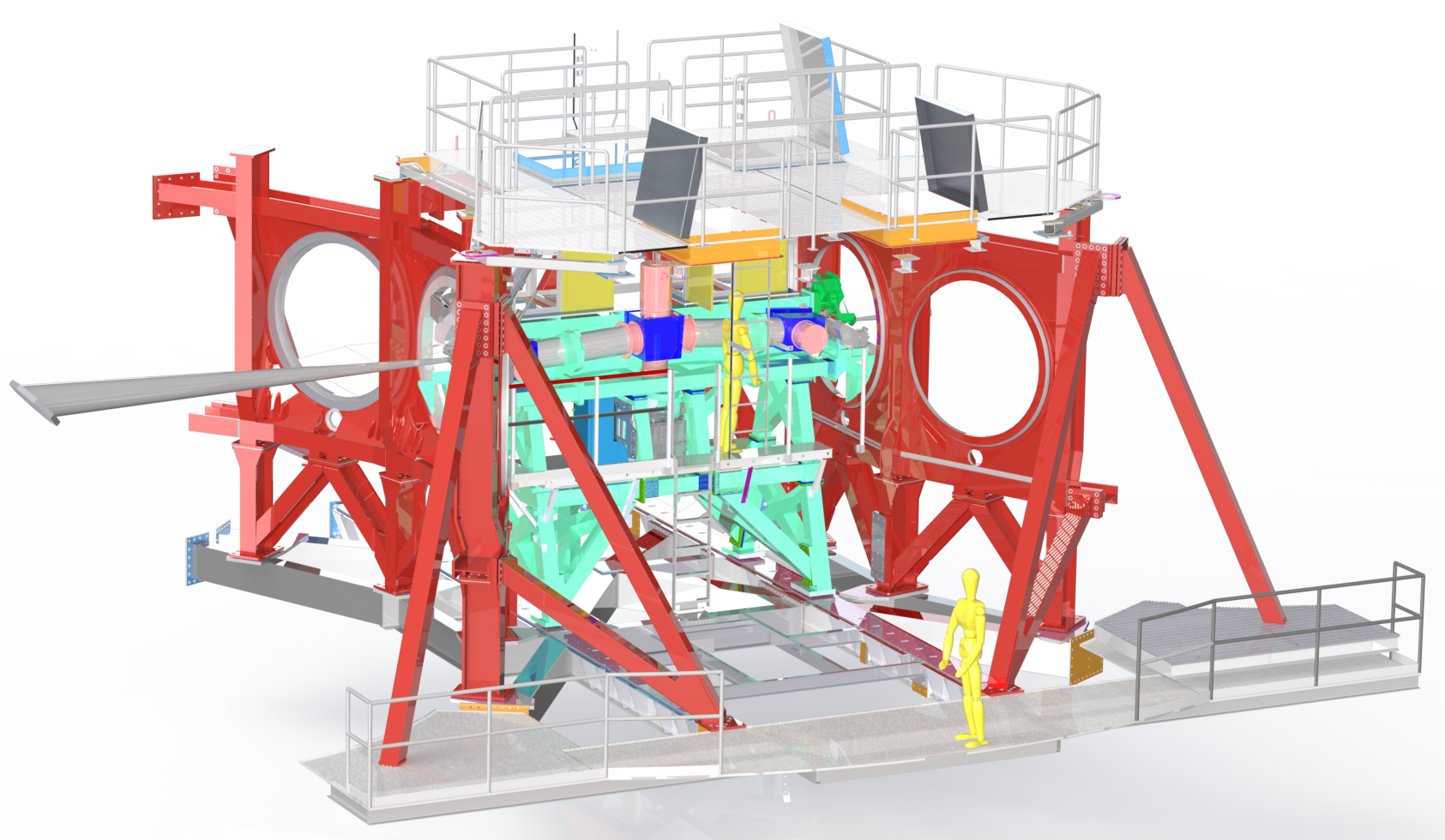} 
        \caption{LBTI (shown in green) within the context of the larger structures of the LBT instrument gallery. This view is from the rear left of the telescope structure.}
    \end{subfigure}
	\begin{subfigure}{\columnwidth}
        \centering
        \includegraphics[width=1.00\textwidth]{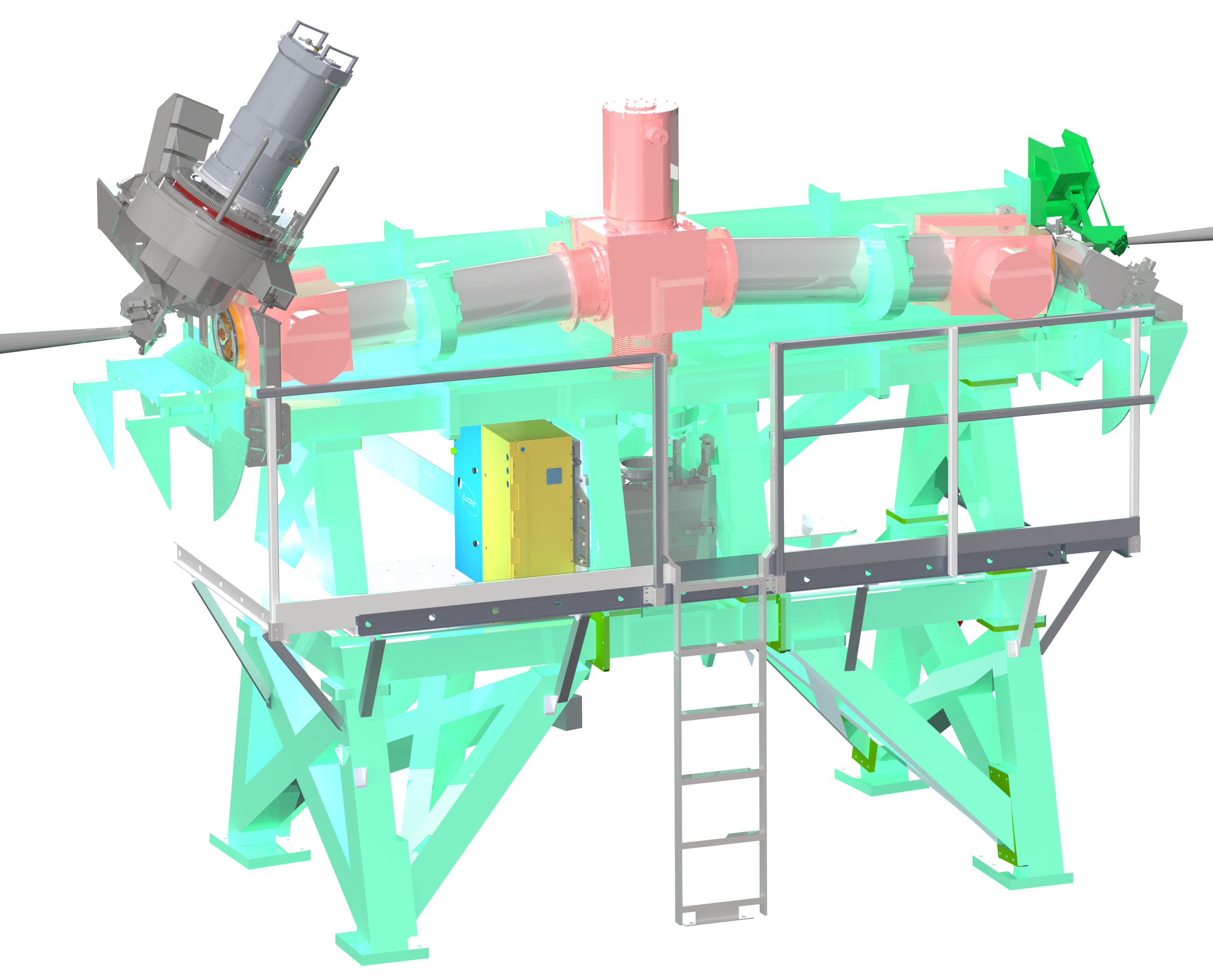} 
        \caption{The iLocater fiber injection system within the LBTI structure.  The SX (blue) and DX (gold) acquisition camera modules are mounted inside one of the instrument bays of LBTI using a custom interface plate to allow vertical and lateral adjustment relative to the LBTI structure.}
    \end{subfigure}
    \caption{Rendering of the iLocater acquisition camera within the context of the LBT instrument gallery and LBTI.}
    \label{fig:lbti}
\end{figure}

The iLocater Acquisition Camera will occupy the SX instrument bay of LBTI. This is a small volume (900$\times$1050$\times$\SI{855}{\milli\meter}) accessible within the LBTI structure (Fig.~\ref{fig:lbti}). The acquisition camera comprises two independent mechanical modules, one for the SX system fed from the left telescope primary, and the other for the DX fed from the right. These modules secure to a custom mounting interface on LBTI which allows for $\pm$\SI{50}{\milli\meter} vertical and lateral adjustment of both modules relative to the incident beam.

The mechanical design of the acquisition camera has been developed around the instrument optical design. An iterative approach was adopted when optical adjustments have been required to accommodate mechanical constraints. The mechanical system has several fundamentals of its design: 

\begin{enumerate}
\item Ensuring stability: to maintain single-mode fiber coupling, the instrument must remain mechanically stable throughout a science observation while the telescope is tracking. As the LBT is a Gregorian design, the instrument orientation will change with telescope elevation and therefore it is important the overall system and components maintain position during these expected changes. This can be achieved by minimizing system flexure in the design and actively holding position of components where necessary to maintain beam position.
\item Modularity: A modular design enables smaller sections of the instrument to be integrated and modified without the need to remove the entire instrument from the telescope. While the entire SX or DX acquisition camera modules can be independently removed, having independent sub-assemblies within each system improves the efficiency of long-term support for the instrument. 
\item Instrument Access: Given the location of the instrument on the telescope, it is desirable to be able to resolve minor hardware issues in-situ with the instrument still mounted to the structure of LBTI.
\item Facilitate Upgrades and Expansion: Having an efficient single-mode fiber injection system enables multiple avenues for technology demonstration and future science upgrades. The mechanical design provides flexibility to enable these.
\end{enumerate}

\subsection{Mechanical Structure}

\begin{figure*}
	\centering
	\begin{subfigure}{0.95\columnwidth}
		\includegraphics[width=\textwidth]{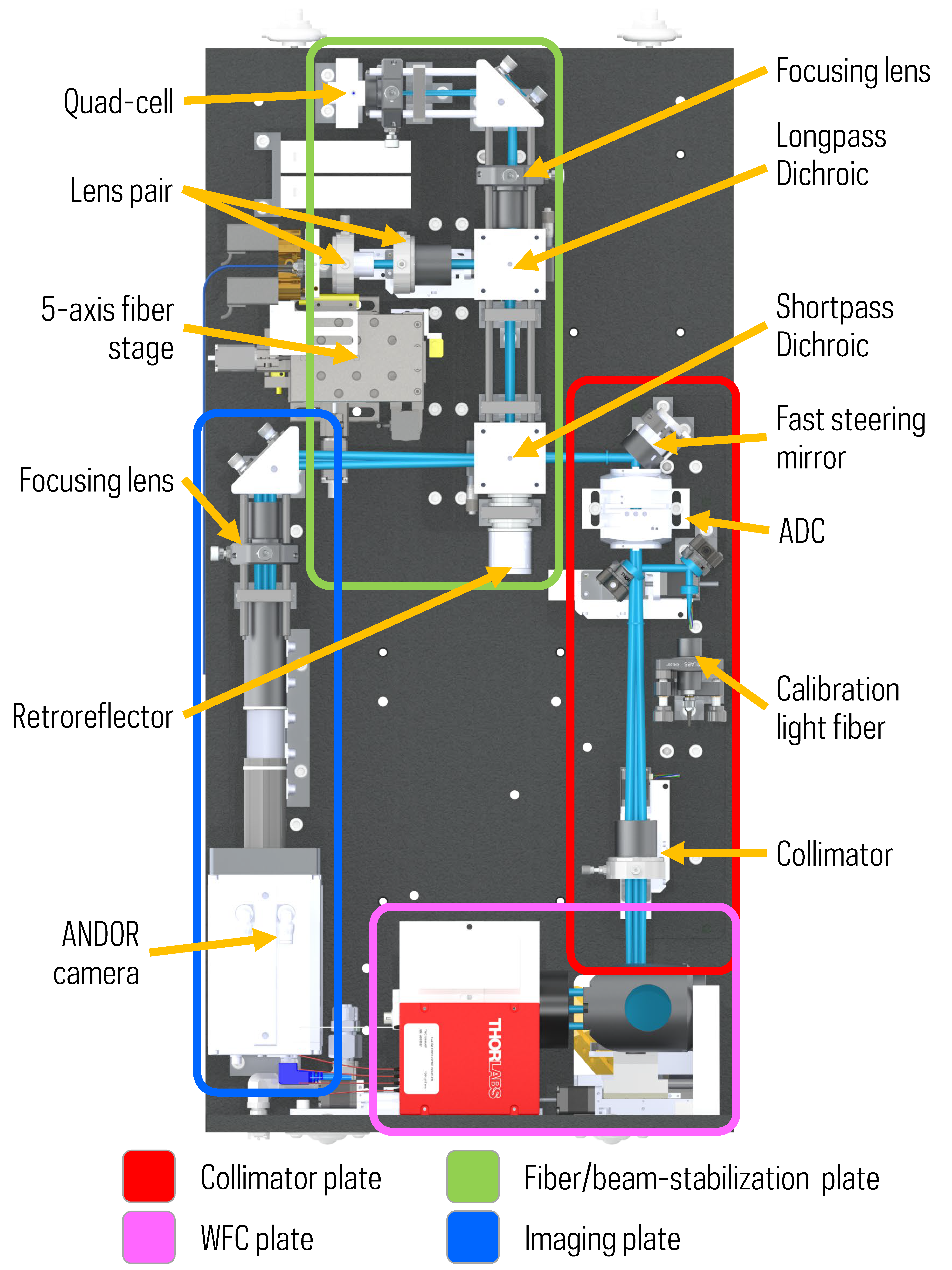}
		\caption{Front view of the mechanical system showing the individual components and sub-section plates mounted onto structural board of the instrument.}
		\label{fig:mechanical_front}
	\end{subfigure}
	\begin{subfigure}{0.95\columnwidth}
		\includegraphics[width=\textwidth]{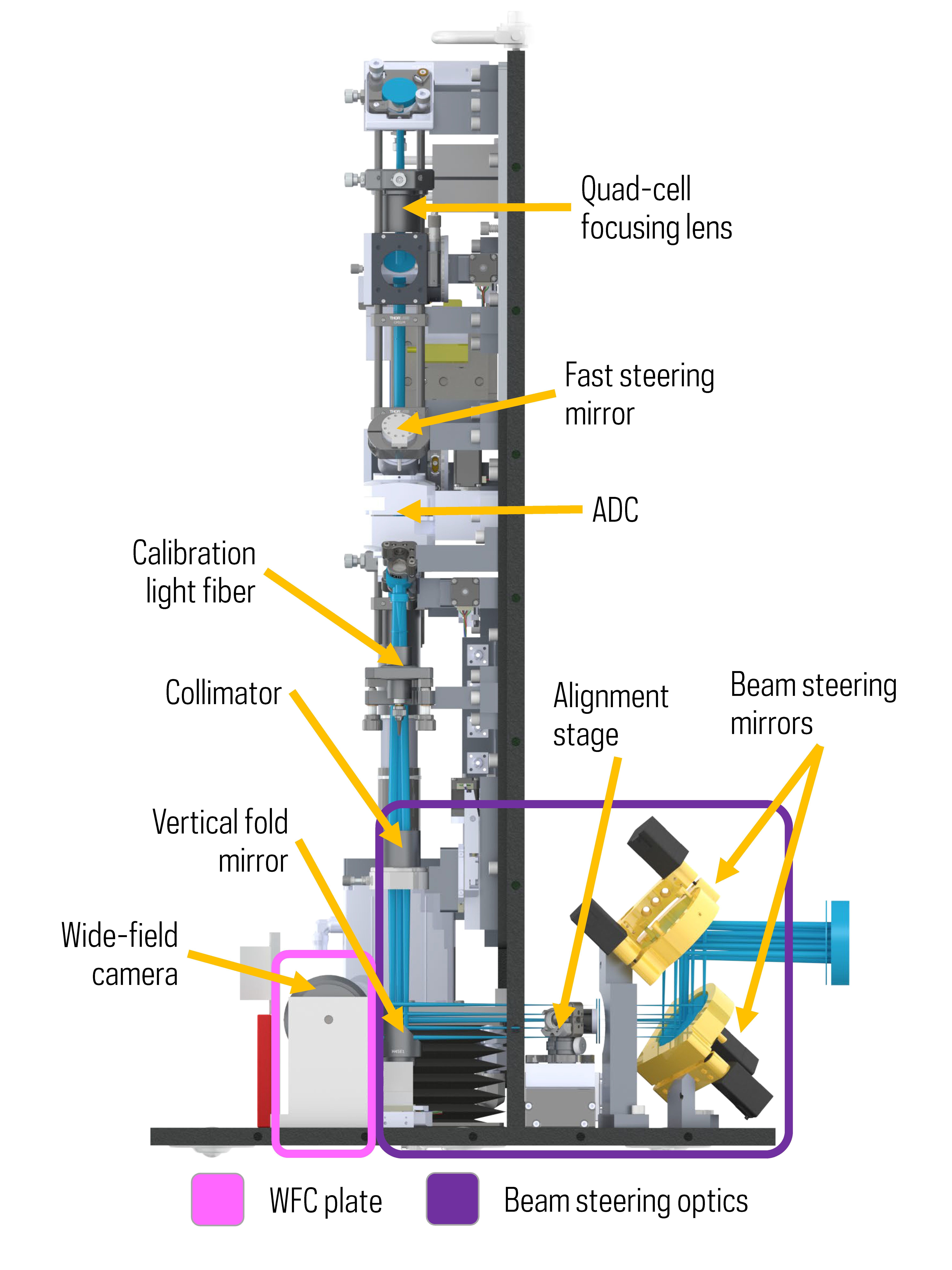}
		\caption{Side view of the mechanical system.}
	\end{subfigure}
	\caption{Renderings of the iLocater SX acquisition camera mechanical system.}
	\label{fig:mechanical_structure}
\end{figure*}

The SX and DX acquisition camera modules are structurally independent of each other. Two 1/2" thick aluminum plates form the major structural support of each mechanical enclosure and these are secured at right angles to each other to effectively form a floor and vertical upright (Fig.~\ref{fig:mechanical_structure}). A sheet metal structure is attached to the plates to increase rigidity and includes features and seals to provide environmental and water ingress protection. The entire enclosure dimensions are 394$\times$488$\times$\SI{768}{\milli\meter} with a weight of \SI{52}{\kilogram}.

The vertical support plate houses the majority of the optomechanical components. These are divided into smaller subgroups, each mounted onto a 1/4" thick aluminum plate and secured to the larger structural boards. Alignment pins and bushings allow the repeatable removal and installation of these modules relative to the main enclosure structure. All optical plates and posts have undergone a hardcoat anodization process including beadblasting where appropriate to minimize the effects of any scattered light within the instrument. A NIR optimized anodization process was used where feasible.

COTS optomechanics and electronics have been used in the instrument where possible to minimize cost and design efforts while also providing required adjustability of components. Modifications to specific components have been made where needed to provide stability, for example, adding locking screws on lateral adjusters in lens mounts. All optomechanics are secured to the instrument plates using single-piece custom machined posts to provide secure and rigid mounting.

The material choice for all the mechanical plates and optical posts was assessed as part of the instrument design and in the context of the previous experience gained with the prototype system installed at the LBT in 2016. The weight and handling of the entire system are of key importance for installation given space constraints (\cref{sec:installation}). These constraints advocated towards a decision to use aluminum, however, its structural properties are not as advantageous for stability as steel. Despite this limitation, through laboratory testing of components, reducing the optical beam-height (\SI{76}{\milli\meter}) and comparing the orientation of the changing gravity vector on the system compared to that of the previously installed system, the performance of aluminum was deemed to be satisfactory particularly when accounting for the active beam-alignment and stabilization capabilities now present within the system (\cref{sec:beam_stability}).

\subsubsection{Beam Steering Opto-mechanics}

A pair of 2" mirrors are used to align the incident beam from LBTI to the instrument optical axis. These are mounted within COTS tip/tilt mirror mounts (Newport U200-AC) which are attached to custom supports with angles machined to position the mirror and mount in nominal position for the optical design. These are mounted directly to the 1/2" base plate of the acquisition camera enclosure. The mirror mounts include stepper motors (Micronix MP-21) to allow remote tip/tilt adjustment of the beam. 

A linear stage (Micronix VT-50) is located after the beam steering mirrors at the nominal focal plane of the f/41.2 beam. Mounted on this stage is a retroreflector, pinhole and fold mirror which are used for instrument integration, internal beam alignment and laboratory testing using a simulated telescope beam which is injected through the removable panel in the enclosure wall. The beam steering system also includes a fold mirror mounted on a linear stage (Micronix VT-21L\footnote{The same model is used for all 1" linear stages in the system.}) which is inserted into the beam to illuminate the common optics.

\subsubsection{Collimator Plate}

\begin{figure}
	\includegraphics[clip, trim=0cm 0cm 0cm 0cm, width=1.0\columnwidth]{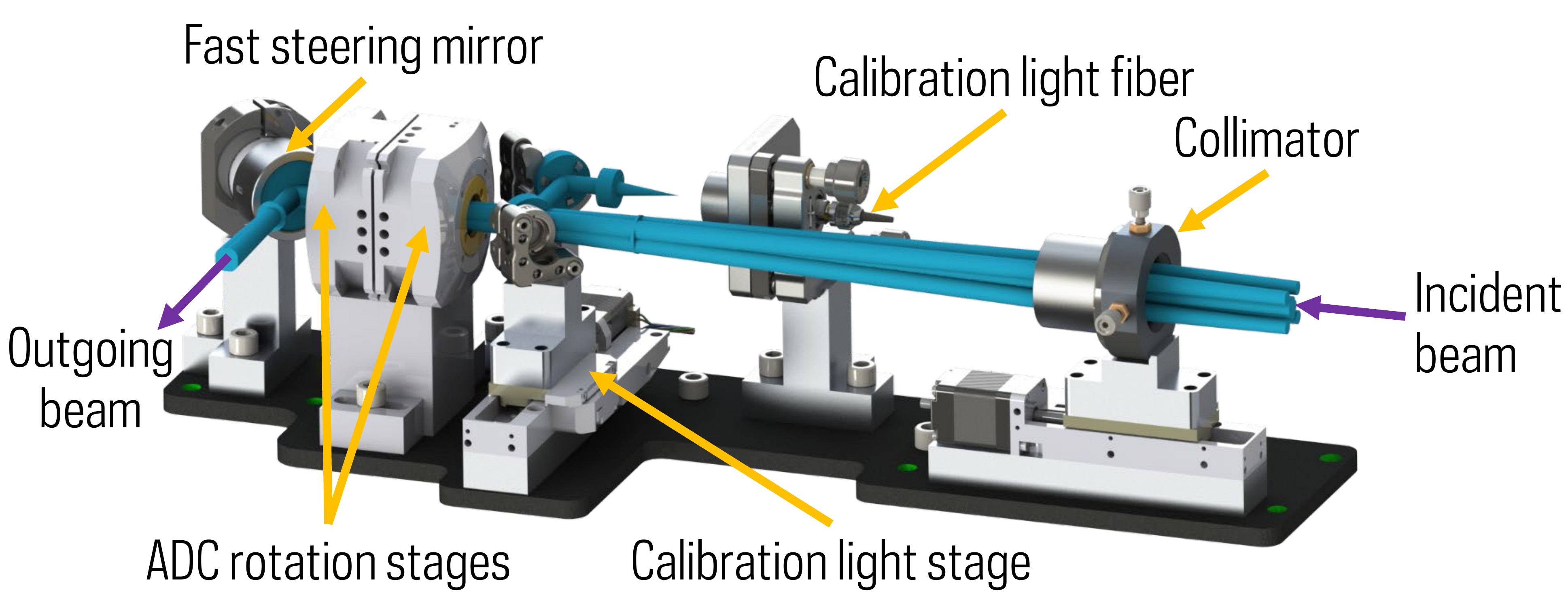}
    \caption{Rendering of the SX acquisition camera collimator plate. The beam from LBTI enters from the
right of the figure, passes through a collimating lens before continuing through the instrument ADC and being
reflected by a fast tip/tilt mirror. Also included is the hardware to inject calibration light to the instrument spectrograph.}
    \label{fig:collimator_plate}
\end{figure}

The collimator plate (Fig.~\ref{fig:collimator_plate}) comprises the f = \SI{257}{\milli\meter} collimator lens mounted on a 1" linear stage, the instrument ADC prisms housed within two counter-rotating mounts and the fast tip/tilt mirror for beam stabilization. The collimator is mounted on a linear stage to allow for internal focus adjustments in addition to the focus adjustment capabilities of the AO system. The lens mount is secured to the stage using a solid spacing block to reduce the beam height and therefore reduce flexure effects. The calibration light channel for the instrument spectrograph is injected into the beam after the collimator using mirrors mounted on another linear stage.

The ADC prisms are mounted in two rotation stages (Newport CONEX-AG-PR100P) using
custom interface adapters. These were developed to allow individual tip/tilt adjustment of each prism and ensuring their alignment to the optical and rotation axes. An alignment procedure including positioning rods and pins has been developed to ensure the rotation axes of the two stages are co-aligned correctly to required optical tolerances.

The tip/tilt mirror in the optical system (nPoint RXY3-276) is mounted in a clamp design to allow adjustment
of the overall mirror position relative to the beam. The clocking of the mirror is controlled using an interface
plate secured to the rear of the mirror stage ensuring the two drive axes are correctly oriented relative to the
incoming beam.

\subsubsection{Fiber Channel \& Beam-stabilization Plate}

\begin{figure}
	\includegraphics[clip, trim=1cm 0cm 0cm 0cm, width=1.0\columnwidth]{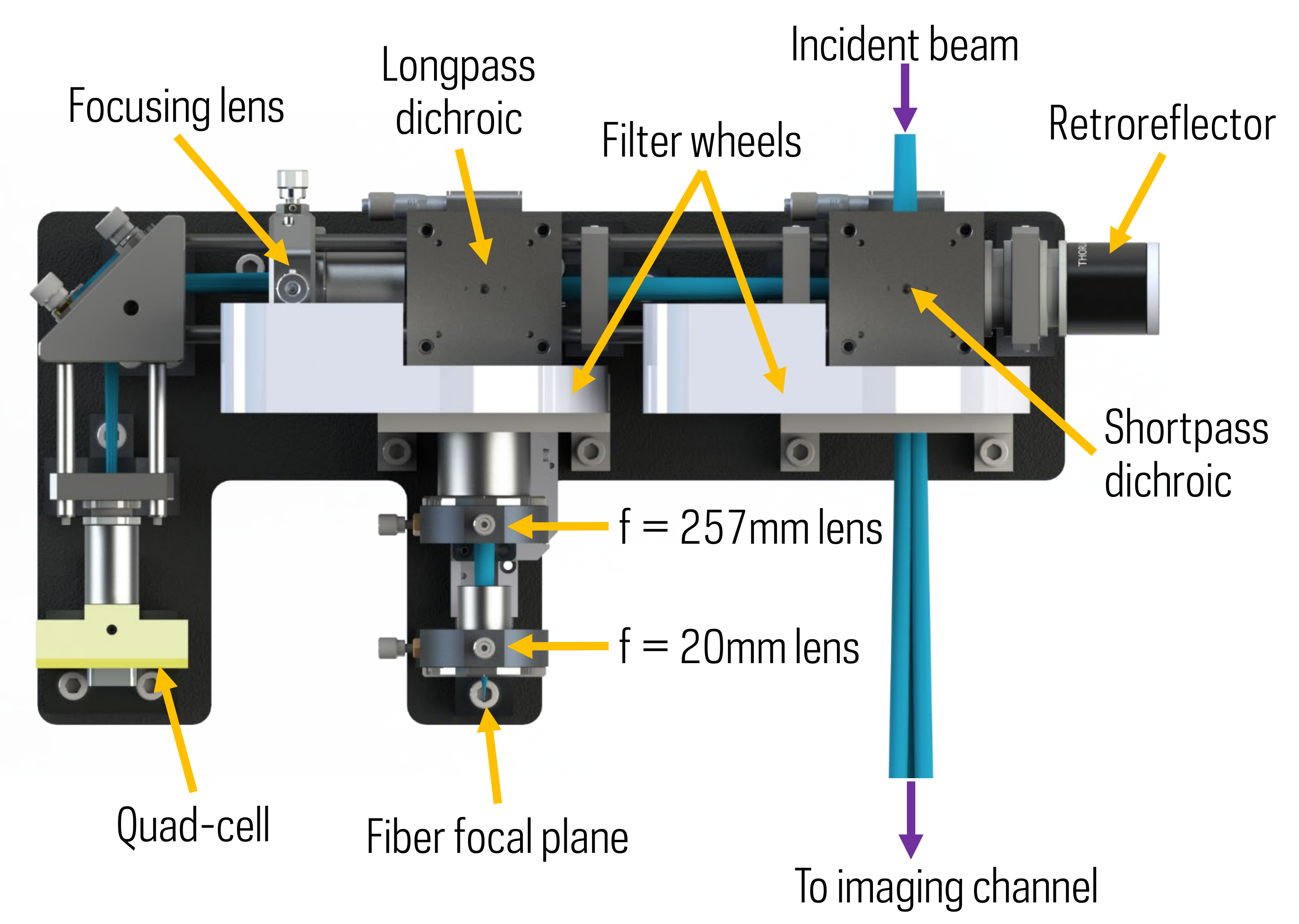}
    \caption{Top down rendering of the cage-based fiber channel \& beam-stabilization plate. The beam from the collimator plate enters from the top right before being split by a dichroic housed within a cage-cube. The short wavelength beam passes directly through to the imaging plate while the longer wavelengths are split again with a second long-pass dichroic to generate the beam-stabilization (transmitted) and fiber (reflected) channels. Each beam is then independently focused.}
    \label{fig:fiber_plate}
\end{figure}

The instrument dichroics are mounted onto a single mechanical plate including the optics for the fiber-coupling and beam-stabilization channels (Fig.~\ref{fig:fiber_plate}). The system is primarily composed of an opto-mechanical cage system which houses the two instrument dichroics on lockable rotation stages to allow precision alignment. Tip/tilt adjustment of these optics is achieved using shimming. The f = \SI{200}{\milli\meter} focusing lens for the beam-stabilization channel and the retroreflector for back-illumination are also mounted within the cage system. 

The pair of lenses in the fiber channel are mounted outside of the cage system to allow precision adjustment. The first lens (f = \SI{257}{\milli\meter}) is mounted on a 1" linear stage to allow adjustment of the beam f/number and spot size in the fiber focal plane. The second lens (f = \SI{20}{\milli\meter}) is mounted in a custom sleeve to ensure co-alignment of the lens to its lens tube. 

The overall optical integration and alignment of the acquisition camera is done by back-propagating beams from the fiber focal plane through the optical system due to the tightest optical tolerances in the system being at this location. Upstream optical modules are aligned with the optical axis defined by the fiber-coupling channel.

The instrument fibers are mounted on a precision three-axis stage (Newport 562F-XYZ-LH) which is secured
directly to the main back-plate of the instrument. This provides improved stability and increased coarse adjustment relative to the beam from the fiber-coupling channel. The stage axes are driven by stepper motors and the mechanical structure has been modified to incorporate encoders to provide absolute position measurements. Mounted on top of the stage is an independent tip-tilt mount (Newport U100-P) also driven by stepper motor actuators giving a total of five degrees of freedom on the fiber position. The range of motion of this stage allows access to three separate FC terminated fibers which can be mounted simultaneously on the stage: the single-mode science fiber, a multi-mode fiber for coupling diagnostics and an expansion port for testing and upgrades.

A pair of filter wheels have been incorporated into the mechanical plate design (one for the fiber channel and one for the imaging channel), however, these have not been included in the constructed system.

\subsubsection{Imaging Plate}

The imaging plate comprises a fold mirror, a focusing lens (f = \SI{257}{\milli\meter}) and the instrument ANDOR camera. The system is primarily constructed of cage-based optomechanics and lens tubes to aid with alignment and stability.

\subsubsection{Wide-field Camera Plate}

The wide-field camera plate comprises a 2" fold mirror followed by a pair of 2" diameter achromatic doublets mounted in a single lens tube to generate the wide field of view required for beam acquisition. A Gigabit Ethernet camera is mounted on a 1" linear stage to allow both focal plane and pupil plane imaging of the incident beam. The lens and linear stage are mounted on an individual sub-plate to allow co-alignment and adjustment relative to each other prior to integration into the full instrument.

\subsection{Enclosure Features \& Interfaces}
\label{sec:installation}

\begin{figure}
	\centering
	\begin{subfigure}{0.95\columnwidth}
		\includegraphics[width=\columnwidth]{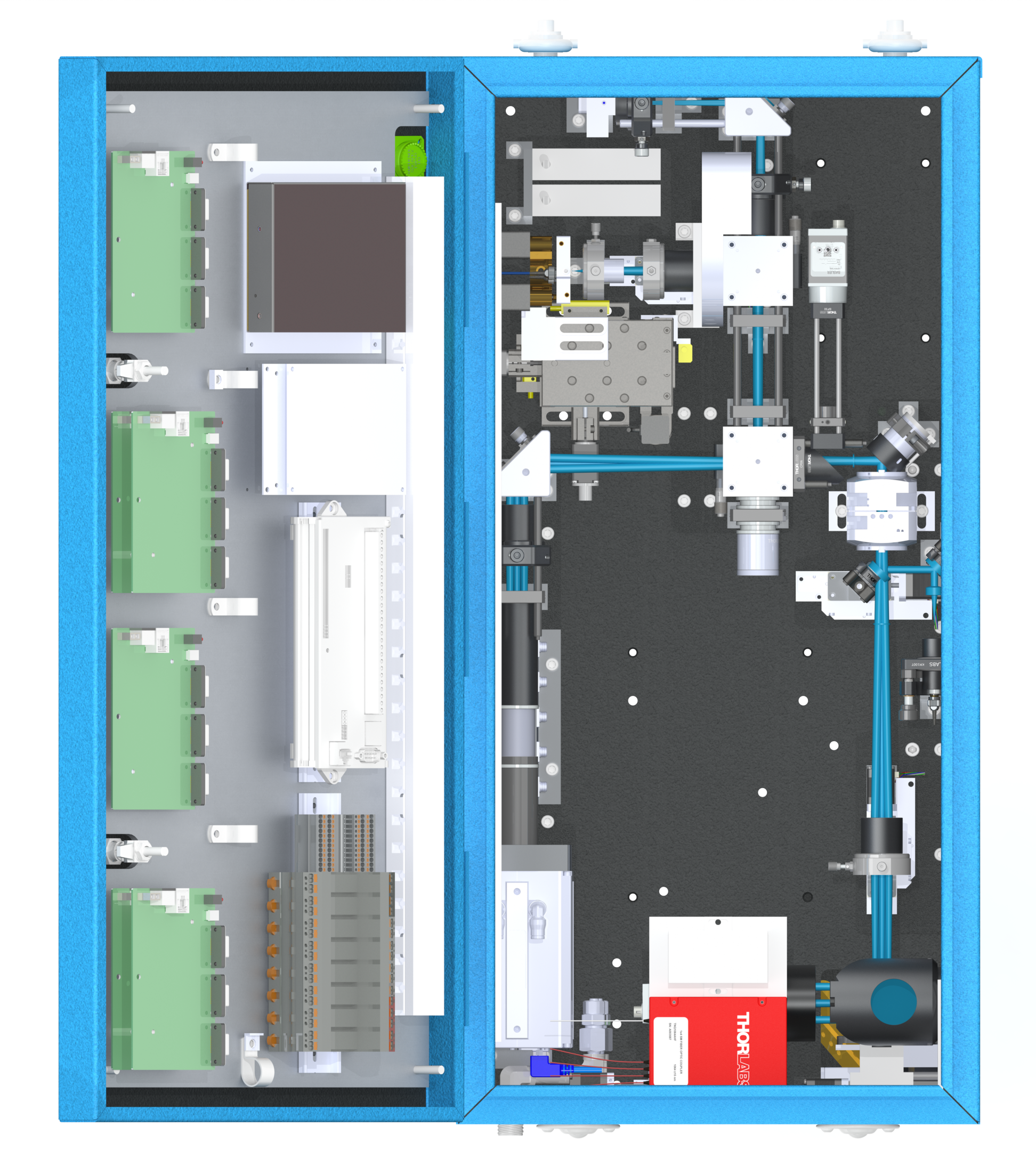}
    	\caption{Rendering of the SX acquisition camera module showing the entire enclosure, electronics plate and connections. Machine rollers and lifting eyes are used during the installation process.}
		\label{fig:enclosure_CAD}
	\end{subfigure}
	\begin{subfigure}{0.95\columnwidth}
		\includegraphics[width=\textwidth]{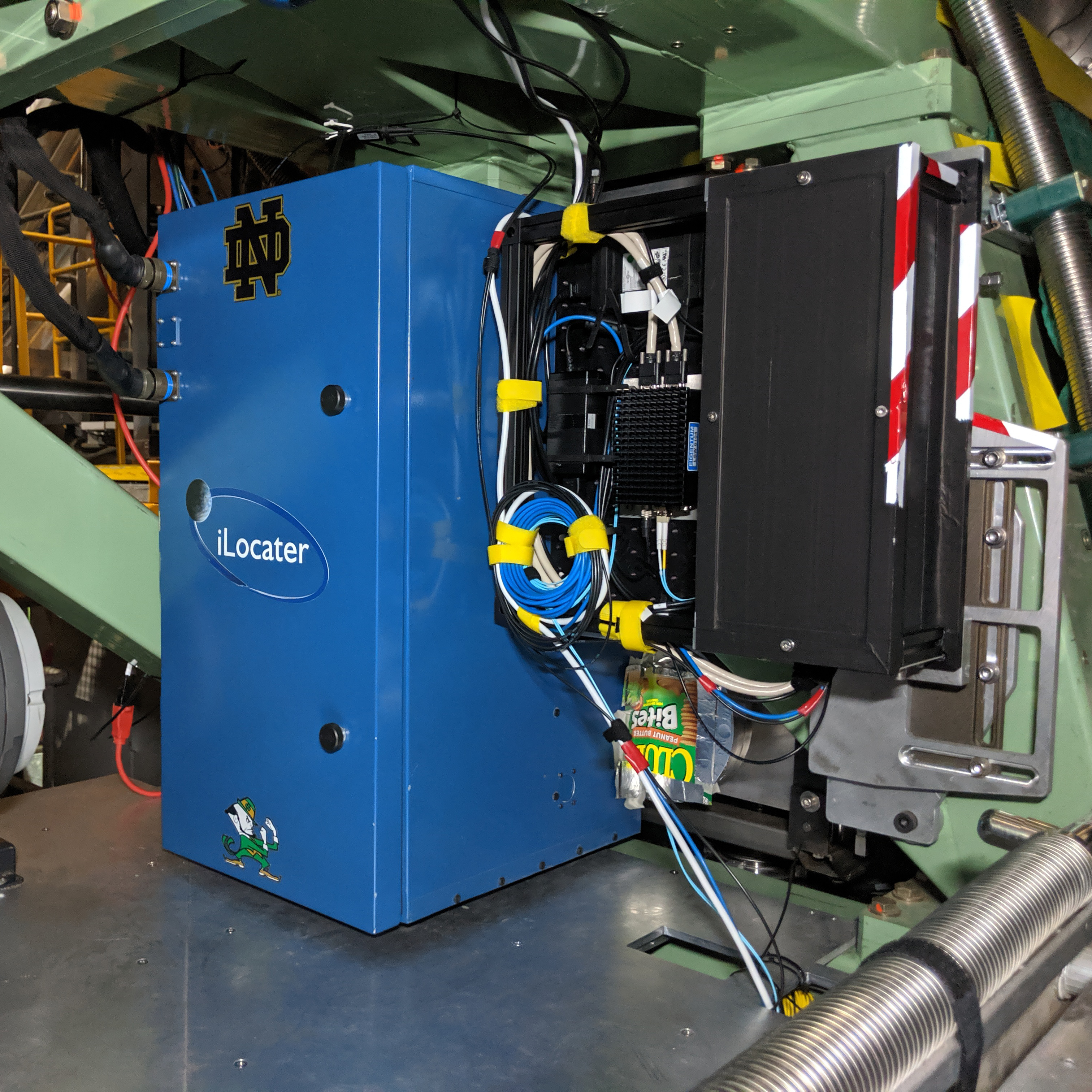}
		\caption{The SX acquisition camera enclosure installed within the instrument bay of LBTI. The sheet metal floor used during installation is visible below the enclosure. Temporary electronics are installed in the location of the DX system.}
		\label{fig:enclosure_photo}
	\end{subfigure}
	\caption{The iLocater SX acquisition camera enclosure.}
	\label{fig:enclosure_full}
\end{figure}   

The mechanical enclosure of the system incorporates handling and installation features for the instrument (Fig.~\ref{fig:enclosure_full}). Lifting eyes are incorporated in the main vertical structure to allow lifting via crane during installation. The base plate of the enclosure contains machine bearings to allow the entire unit to be smoothly rolled across a flat surface. A sheet metal aluminum floor has been installed within the SX instrument bay of LBTI to allow for the instrument transfer into position (Fig.~\ref{fig:enclosure_photo}).

Where possible, instrument electronics are located remotely from the acquisition camera in a thermally isolated rack, however, specific components with cable length constraints are housed within a small electronics bay inside the enclosure door. The electronics volume is separated from the optical components by removable 1/4" thick polycarbonate panels which are designed to prevent any light and thermal leakage from the installed electronics.

All interfaces into the enclosure are designed to be watertight where possible and have been situated in optimal locations for internal cabling and connections. The primary electrical and data connections to the enclosure are situated on the door behind the electrical compartment. A secondary set of connections are situated on the base of the enclosure including chilled water, data-fiber and optical fibers (science and illumination sources). Having a single location for all fiber connections below the unit minimizes the risk of damage when the unit is installed while still allowing efficient installation and removal. A square 4" removable panel is included behind the fiber stage to allow for optical alignment as well as any required diagnostic connections.

\section{Control \& Data}
\label{sec:control}

The iLocater instrument control system has been designed to integrate all sub-systems of the instrument including the acquisition camera, spectrograph and wavelength calibration system. The software is a multi-layered architecture to facilitate decoupling of hardware and software components. This allows easier maintenance of the codebase and greater testing and validation capabilities. Overall communication is handled using the Instrument Neutral Distributed Interface (INDI) protocol which is also used for LBTI control \citep{Downey2007}. This allows effective communication, monitoring and control between iLocater and LBTI. Instrument control is provided through a web interface, command line and scripting protocols. A sub-set of the system has been commissioned for use with the SX acquisition camera. All hardware within the acquisition camera supports a standard set of protocols (USB, RS232, RS485, Ethernet, or remote analog/digital I/O) to aid the software development and integration. 

\subsection{Optomechanics mechanisms}

For simplicity and to aid with development and integration, where possible identical linear stages or stepper motor actuators have been used within the optomechanical system. Components from Micronix were selected which use identical control boards to provide spare capacity and redundancy. Alternative hardware was used for the ADC rotation stages, tip/tilt mirror for beam stability (\cref{sec:beam_stability}) and the fiber stage where specific needs and loading considerations precluded standardization of hardware. All hardware was assessed for suitability for the telescope environment including mechanical stability and performance over the expected annual temperature range (\SIrange[retain-explicit-plus]{-20}{+25}{\celsius}).

All linear stages in the optical path include direct encoder feedback. This allows repeatable positioning of components relative to a known optimal position and closed-loop position holding. Testing of mechanisms in the optical path where encoders are not incorporated, for example tip/tilt mounts, was undertaken and stability was shown to be adequate to hold alignment with changing orientation.

\subsubsection{ADC Control}
\label{sec:adc_control}
In order to ensure good atmospheric dispersion correction throughout an observation, the ADC rotation stage positions are updated based on telescope altitude. As the LBT is an alt-azimuth design, the direction of dispersion as seen by the acquisition camera is static (although field rotation occurs) and therefore only the amount of counter-rotation between the ADC prisms and not their combined orientation needs to be adjusted for altitude.

Using the Zemax optical model, an exported lookup table of optimized prism separation was generated for telescope elevation in steps of \SI{0.5}{\degree} within the operating range (\SIrange{5}{60}{\degree}). A Python script combined with INDI queries the telescope elevation to match this to the nearest value in the generated lookup table before commanding the ADC rotation stages. A manual operation mode is also included for diagnostic and development purposes.

\subsection{Fiber Positioning}
\label{sec:fiber_positioning}

For successful operation, the acquisition camera must allow precise positioning of the science fiber to the incident optical beam. To allow for diagnostics and expansion, it must also allow two additional fibers (FC terminated) to be placed in the beam. This places range and precision constraints on the fiber positioning system.

\subsubsection{Fiber Stage}

To achieve these requirements, upgrades have been made to the same fiber positioning system employed in the demonstration system installed at the telescope in 2016 (Newport 562F-XYZ-LH). While piezo actuators were used previously for their precision, they were not accurate or repeatable and required a complex electronics interface. To resolve these issues, the design of the stage was modified to incorporate stepper motors (Hayden Kerk P21H4U) that were compatible with the standard Micronix control boards used for other mechanisms. Optical encoders from Renishaw (V2BCH20A01F) with \SI{50}{\nano\meter} step size were also added to provide accurate and repeatable positioning. By orienting the three fibers in an optimized triangle configuration, this solution provides sufficient travel in all three fiber stage axes while delivering sufficient repeatability to switch between fibers without the need for any additional coarse acquisition process.

Fig.~\ref{fig:mechanical_front} shows the orientation of the SX acquisition camera when the telescope is at zenith. As the telescope slews towards horizon, the system rotates in the counter-clockwise direction. To ensure sustained contact between the motors and stage, the fiber stage has been orientated to ensure the gravity vector always has a component onto the actuators. This was found to be a limitation of the previous system. By using encoders and motors in closed-loop, the system is able to hold closed-loop position to better than \SI{150}{\nano\meter} over the entire telescope elevation range. For a typical telescope track during an observation, fiber position is maintained to better than \SI{100}{\nano\meter}.

\subsubsection{Fiber Coupling Procedure}
\label{sec:fiber_coupling}
Efficiently injecting the incident telescope beam into the instrument fiber requires a coarse acquisition of the beam, optimization of the beam profile using the telescope AO system, and then optimal positioning of the fiber relative to the beam. The process is iterative to achieve optimal coupling and is outlined below. Additional diagnostic hardware located off the telescope is used in this process to assess coupling (\cref{sec:diagnostic_data}).

\begin{figure}
	\centering
	\includegraphics[width=0.75\columnwidth]{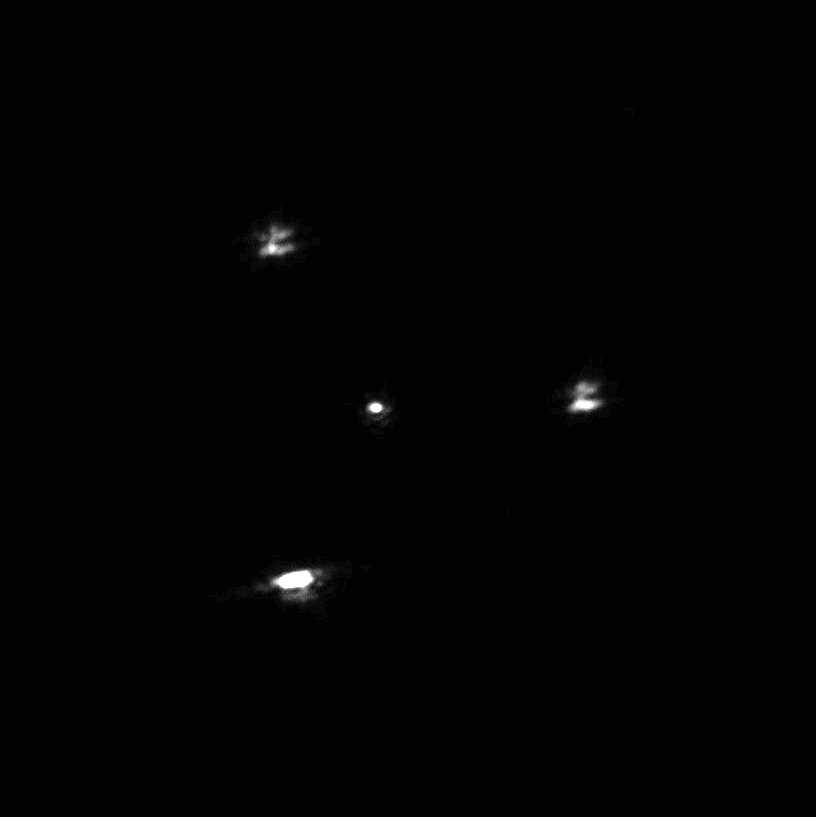}
    \caption{Image of on-sky fiber coupling of HD 200253, (V=6.0, G type) within the imaging channel. The target star is positioned in the center of the three surrounding beams by adjusting the fiber stage position. The three reference beams are generated by back-illuminating from the fiber focal plane. Aberrations are present on the reference images due to effects of the retroreflector on the beam. The nominal distance between the star and off-axis images is $\sim$\ang{;;0.8}.}
    \label{fig:fiber_coupling}
\end{figure}

\begin{enumerate}
	\item Close and stabilize the telescope AO loop.
	\item Acquire the incident beam on the WFC and ensure pupil and focal plane images are in nominal position. Adjust the beam position using a combination of the AO system, UBC optics, and beam steering mirrors to achieve the nominal performance.
	\item Switch to imaging using the ANDOR camera in the imaging channel. Enable ADC correction and adjust focus using the AO system or collimator lens position.
	\item Enable the beam-stabilization loop (\cref{sec:beam_stability}) to remove residual tip/tilt from telescope vibrations.
	\item Inspect the PSF quality on the ANDOR camera. If aberrations are noted, perform NCPA correction using the AO system.
	\item Begin back-illumination of off-axis fibers to generate alignment beams on the ANDOR camera (Fig.~\ref{fig:fiber_coupling}). Adjust the fiber stage position to center the incident beam correctly between reference spots.
	\item When the fiber is nominally positioned, check for a signal on FemtoWatt Receiver (FWR)  (\cref{sec:diagnostic_data}). If necessary, raster scan the fiber stage until signal is acquired. Once a signal is obtained, disable back-illumination.
	\item Continue to raster scan fiber stage in successive axes to converge on optimal position for coupling as determined by signal on the FWR.
	\item If required, reoptimize NCPA correction to check for improvements in coupling.
	\item Engage closed-loop operation of the fiber stage to hold position during telescope tracking.
\end{enumerate}

\subsection{Beam Stability}
\label{sec:beam_stability}

Extensive work has previously been undertaken to understand the vibrations characteristics of the LBT using the Optical path difference and Vibration Monitoring System (OVMS+), an array of accelerometers mounted on the telescope structure \citep{Kuerster2010, Boehm2016, Escarate2017}. Work in \citet{Escarate2018} demonstrated significant vibration contributions being imprinted onto instrument PSFs during AO operations. Over 90\% of residual jitter amplitude measured by OVMS+ was concentrated into two frequency bands \SIrange{0}{30}{\Hz} and \SIrange{50}{70}{\Hz}, however, some frequencies noted in the PSF jitter were not recorded by OVMS+. Therefore, despite plans to implement a telescope wide feed-forward correction using OVMS+, residual tip/tilt errors will still be present in the beam. Given the stringent stability requirements of fiber coupling, an internal closed-loop beam-stabilization system has been implemented within the acquisition camera system targeting the identified vibration bands.

The beam-stabilization loop is based on a Hamamatsu G6849-01 InGaAs quadrant detector (quad-cell) and associated electronics. This was initially developed by NASA JPL and has been modified for specific use within iLocater. The quad-cell outputs a differential voltage for horizontal and vertical position as well as a summed voltage. The relative position of the beam on the quad-cell device is determined by dividing the differential signals by the summed signal. This processing takes place on a custom analog divider board developed at Notre Dame with independent horizontal and vertical outputs providing proportional corrective signals for the system tip/tilt mirror. The internal PID loop of the mirror controller (nPoint LC.402) is adjusted to provide optimal correction. The correction bandwidth is limited by the quad-cell detector electronics and PID process within the mirror controller. Optimum correction is realized within the \SIrange{0}{30}{\Hz} vibration band, with diminishing performance into the \SIrange{50}{70}{\Hz} band. This is sufficient to correct for the highest amplitude vibrations recorded at the LBT. 

\subsection{Diagnostic Data}
\label{sec:diagnostic_data}

To aid with assessing instrument performance, several datasets were recorded or monitored as part of the acquisition camera operation. These include fiber throughput data recorded using a FWR, high-frame rate imaging of the PSF using the ANDOR camera at the imaging channel focal plane, real-time AO system residuals, and vibration data from the telescope OVMS+ system.

\subsubsection{Fiber Throughput Data}

The FWR is an optical system which has been designed to measure the flux output from optical fibers. The system is illuminated using an FC terminated multi-mode bifurcated fiber (FG105LCA) with two inputs. This beam is collimated, passes through a pair of filter wheels, and is refocused onto a single high gain transimpedance amplifier coupled photodiode (Thorlabs PDF10C) which is read out using an oscilloscope. The photodiode has a bandwidth of \SI{25}{\Hz}. Using common optics for the optical paths for the two fibers minimizes calibration errors and allows a direct comparison of power between the two fibers. A pair of filter wheels are included in the collimated space of the system, one containing neutral-density filters and the other containing  narrow-band filters spanning the iLocater spectrograph bandpass. This allows broadband fiber throughput to be assessed over a dynamic range spanning the full expected target magnitudes for iLocater.

The fiber inputs of the FWR are connected via patch fibers to the SMF and MMF (FG105LCA) installed at the fiber focal plane in the acquisition camera. The MMF at the focal plane has a core diameter of \SI{105}{\micro\meter}, over 12 times larger than the SMF MFD at the longest fiber coupling wavelength and well in excess of the PSF diameter. This ensures all incident light at the fiber focal plane is injected into the MMF.

For laboratory testing, short patch fibers were used to connect the fibers at the focal plane to the FWR, however, at the telescope the FWR system was installed remotely from the acquisition camera system in the location which will ultimately be occupied by the instrument spectrograph. The science fiber was connected to the FWR using a \SI{40}{\meter} AVIM terminated SMF (manufactured from the same fiber batch selected for iLocater) combined with a AVIM to FC/PC patch cable and short MMF patch cable. The \SI{40}{\meter} SMF will ultimately be used to feed the instrument spectrograph. The MMF in the fiber focal plane was connected using an FC/PC terminated \SI{40}{\meter} MMF patch fiber combined with a second short MMF patch cable. The end-to-end throughput of the SMF and MMF systems was calibrated by illuminating each fiber in the acquisition camera with a stable laser source and measuring the flux at each break along the fiber chain using a power meter. Combining this throughput with a calibration of the two fiber inputs and internal systematics of the FWR, this system allows a direct measurement of the total flux at the fiber focal plane using the MMF and flux output from the SMF. By taking the ratio of the two calibrated measurements, this provides the coupling efficiency of the SMF. 

\subsubsection{PSF Imaging}

The ANDOR camera in the imaging channel provides high-frame rate imaging of the target star. Full frame read-out (2048$\times$2048) can be achieved at \SI{40}{fps} while smaller regions can be read out at higher speed (e.g. a 128$\times$128 window can be recorded at \SI{1627}{fps}). The acquisition camera control hardware is able to record short periods of data from this camera which allows monitoring and characterization of both AO performance via PSF characteristics and any residual tip/tilt on the incident beam by assessing PSF centroid location.

\subsubsection{Telescope Diagnostics}

In addition to the internal diagnostics of iLocater, the telescope AO system and OVMS system provide further information. While the AO system does not extensively log performance characteristics due to the volume of data this would generate, residual wavefront error is available as part of the control loop and this was monitored for performance during on-sky operation. Additionally, the live frequency spectrum from OVMS is available and can be monitored for any particular vibration peaks which may be impacting performance.

\section{Laboratory Testing}
\label{sec:labperf}

To assess system performance during integration, a beam simulating the incident f/41.2 beam from LBTI was injected into the acquisition camera optical system. This was achieved by removing a small cover on the side of the instrument enclosure and inserting the fold mirror mounted on the alignment linear stage next to the beam steering mirrors. The f/41.2 beam is generated by a `telescope simulator' which comprises a pair off-axis parabolic (OAP) mirrors, one collimating and one focusing, with an iris situated between them which is used to truncate an oversized beam and generate a flat wavefront with the appropriate beam characteristics. An SMF is used to illuminate the system which enables sources to be used interchangeably. A pair of adjustable fold mirrors (one with fast tip/tilt control) are located after the second OAP to allow positioning of the simulator beam onto the instrument optical axis.

Characterization of both the laboratory and on-sky system performance has primarily utilized two measures: the point spread function (PSF) properties recorded on the ANDOR camera in the imaging channel and fiber coupling efficiency measured by the FWR. To assess imaging performance, a 2D Gaussian least-squares fitting algorithm has been used to determine the PSF profile and characteristics. These values are then assessed relative to the nominal optical design. During laboratory testing, imaging quality was also assessed in the fiber focal plane using a temporarily installed detector and the same fitting process.

\subsection{Imaging Channel Performance}

The PSF within the imaging channel was assessed relative to the expected parameters derived from the Zemax optical model of the system. Broadband illumination of the telescope simulator was used to generate the incident beam. The mean FWHM of the PSF with 1$\sigma$ confidence interval in the imaging channel was measured as $6.28\substack{+0.49 \\ -0.09}$ pixels. The expected FWHM at \SI{950}{\nano\meter} from the optical model is 6.1 pixels. The estimated maximum measured Strehl ratio for the system when fed using the telescope simulator exceeds 90\%.

\subsection{Fiber Channel Performance}

Imaging of the PSF directly within the fiber channel is challenging due to the small diameter needed for efficient fiber coupling and typical detector pixel sizes. The nominal FWHM of the PSF based on the optical design is \SI{3.7}{\micro\meter} at a wavelength of \SI{1}{\micro\meter}. To measure the PSF, the system was illuminated using a source at \SI{1050}{\nano\meter} injected through the telescope simulator and imaged onto a \SI{1.6}{\micro\meter} pixel CMOS detector (Basler acA3800-14um) situated at the fiber focal plane. A value of 3.5$\pm$\SI{0.4}{\micro\meter} was found, consistent with the expected performance.

SMF coupling efficiency was also measured as a further verification of performance in the fiber channel. A coupling efficiency of 74$\pm$1\% was achieved when comparing the throughput of a coupled SMF and MMFs at the fiber focal plane. This is within a few percent of the maximum theoretical coupling efficiency (78\%) when illuminating a SMF with an Airy pattern.

\subsubsection{Fiber Channel Throughput}

Using a stable source to illuminate the telescope simulator and a portable power meter, the optical throughput of the system was measured between the front surface of the the collimator in the common optics to the fiber focal plane. The total throughput was found to be 91$\pm$1\% with the largest source of loss (5$\pm$1\%) being the COTS gold-coated tip/tilt mirror. Assuming a similar loss at each mirror within the system, this gives a current end-to-end throughput of 81\% from the entrance of the acquisition camera to the fiber focal plane. Throughput will be increased from this value with the planned replacement of the four COTS mirrors with multi-layer coated equivalents. With this performance, the throughput from sky to the fiber focal plane is expected to be $\sim$50\%.

\subsection{Beam Stabilization Performance}

Data from the prototype system installed at the LBT in 2016 was used to assess beam stabilization performance. High speed imaging data of the on-sky PSF had been recorded with this system for a range of targets and conditions. By extracting the PSF centroid location from these images, the same motion was able to be injected into the telescope simulator beam using its tip/tilt controllable mirror, replicating the frequency spectrum of vibrations seen at the telescope. This system was able to emulate telescope vibrations up to \SI{100}{\Hz}.

Beam stabilization performance was optimized by tuning the PID values in the nPoint mirror controller. Given the dynamic nature of the factors which impact the characteristics of the beam (e.g. target brightness, AO system performance, telescope vibrations), PID tuning and stability optimization was done manually for each configuration rather than relying on standard values. Testing included adjusting illumination brightnesses to simulate differing magnitudes, playback of different datasets from the prototype system, and adjusting the amplitude of playback. Without any correction, when playback was enabled, coupling efficiency dropped from greater than 70\% to $\sim$40\% with a typical scatter of 10\%. In closed-loop operation, coupling efficiency increased to $\sim$60\% and scatter was reduced to 5\%. This is consistent with the PSF centroid position noted in the imaging channel where during closed-loop operation, the range and standard deviation of the centroid from the median position was reduced by a factor of two. For a well optimized loop, the standard deviation of centroid motion was recorded as 1.6 pixels in the imaging channel, a small fraction of the PSF width. Through additional tests of the system using a sine-wave error input with varying frequencies, the remaining beam motion and coupling losses and are attributable to higher frequency vibrations which are beyond the correction bandwidth of the system.

\section{Telescope Testing}
\label{sec:skyperf}

The SX acquisition camera was delivered to the LBT in June 2019. Reassembly, performance verification, and installation of the system onto the telescope gallery took place over a period of three weeks which was followed by day-time on-telescope optical testing. On-sky commissioning took place during two sets of observing runs, each consisting of three half-nights (8-10 July 2019 and 10-12 November 2019). Two half nights of commissioning were lost to weather. Since on-sky commissioning, a limited set of remote testing has been completed to begin assessing longer-term instrument performance.

\begin{table*}
\begin{tabular}{lllll}
\hline
\hline
\textbf{Target}         & \textbf{Magnitude}                                                   & \textbf{Spectral Type} & \textbf{Seeing}           & \textbf{Testing Metric}                                                                       \\ \hline \hline
HR~6923        & V = 5.05 \& 8.07                                            & A1V \& F5V    & \ang{;;1.0}             & Imaging plate scale                                                                     \\ \hline
2 Cygni        & \begin{tabular}[c]{@{}l@{}}V = 4.98\\ J = 5.15\end{tabular} & B3IV          & \ang{;;0.65}            & \begin{tabular}[c]{@{}l@{}}ADC assessment (imaging)\\ Beam Stabilization (imaging)\end{tabular} \\ \hline
HD~200253      & J = 4.68                                                    & G5III         & \ang{;;1.0}             & Beam Stabilization  (fiber coupling)                                        \\ \hline
HIP~28634      & V = 6.49                                                    & F5            & \ang{;;0.9}             & Extended duration fiber coupling (20 mins)                                              \\ \hline
UCAC2~30573218 & R = 13.50                                                   & M4            & \ang{;;1.28}            & Faint target fiber coupling                                                                    \\ \hline
HR~5553        & \begin{tabular}[c]{@{}l@{}}R = 5.50\\ J = 4.88\end{tabular} & K0.5V         & \ang{;;1.33} - Variable & Narrowband fiber coupling                                                        \\ \hline
\end{tabular}
\caption{The primary targets used for on-sky instrument and fiber coupling performance assessment.}
\label{tab:targets}
\end{table*}

\begin{figure}
    \centering
    \includegraphics[clip, trim=0cm 0cm 0cm 0cm, width=0.8\columnwidth]{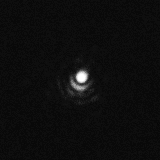}
    \caption{A frame from the acquisition camera ANDOR camera showing an NCPA corrected PSF for the target HR8799. The image bandpass is \SIrange{0.92}{0.95}{\micro\meter} with an estimated Strehl ratio of 52\%. The first diffraction minimum is visible at an angular distance of $\sim$30mas from the center of the PSF core.}
    \label{fig:optimized_psf}
\end{figure}

On-sky performance of the system has been characterized using targets spanning a range of spectral types and magnitudes (Table~\ref{tab:targets}). Tests performed included assessing ADC and beam-stabilization performance and to identify any future upgrades and modifications required for optimal fiber coupling. All on-sky observations were made with the newly installed SOUL AO system hardware on LBTI. During commissioning observations, the AO hardware was still undergoing optimization to reach its full performance. NCPA correction for each observations was assessed visually using the PSF in the imaging channel. Individual Zernike coefficients in the AO system correction were tuned for each target to remove visual aberrations and optimize the peak pixel value recorded on the ANDOR camera. A nominal set of coefficients for NCPA correction was obtained during day-time testing prior to on-sky observations. An optimized on-sky PSF from the AO system is shown in Fig.~\ref{fig:optimized_psf}.

\subsection{Imaging Channel Characterization}

Optical characterization of the imaging channel was completed with observations of the resolved binary system HR~6923. The two system components (separation$=$\ang{;;3.8}) were recorded using the ANDOR camera in a series of short (\SI{0.1}{\sec}) exposures \citep[WDS;][]{Mason2001}. In each frame the centroid location of the components was calculated using a 2D Gaussian least-squares fitting algorithm. The derived separation of the components in each frame was median-averaged to determine the measured plate scale of 0.608$\pm$\SI{0.08}{\arcsec\per\milli\meter}, matching the expected design performance of \SI{0.607}{\arcsec\per\milli\meter}. The calculated field of view of the imaging camera derived from the plate scale is 8.09$\times$\ang{;;8.09}.

\subsection{ADC Performance}

ADC performance has been characterized by measuring the FWHM of the PSF in the imaging channel and assessing any elongation due to dispersion. The imaging channel was selected for characterization as it decouples the PSF profile from centroid motion which would impact fiber coupling observations. Given the limited commissioning time available, it was determined that the assessment of the PSF profile was the most efficient method for characterization despite the imaging channel having limited bandpass (\SIrange{0.92}{0.95}{\micro\metre}) and therefore having reduced sensitivity to dispersion.

\begin{figure}
    \centering
    \begin{subfigure}{\columnwidth}
        \centering
        \includegraphics[clip, trim=2cm 1cm 2.5cm 1cm, width=0.89\textwidth]{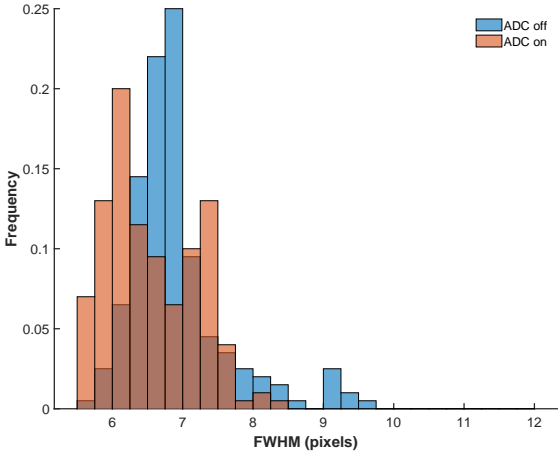}
        \caption{Normalized histogram of PSF FWHM with (orange) and without (blue) atmospheric dispersion correction enabled in the X-axis of the ANDOR camera.}
    \end{subfigure}
	\begin{subfigure}{\columnwidth}
        \centering
        \includegraphics[clip, trim=2cm 1cm 2.5cm 1cm, width=0.89\textwidth]{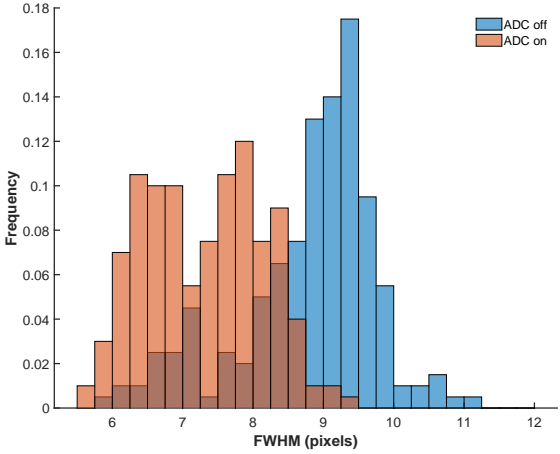} 
        \caption{Normalized histogram of PSF FWHM with (orange) and without (blue) atmospheric dispersion correction enabled in the Y-axis of the ANDOR camera.}
    \end{subfigure}
    \begin{subfigure}{\columnwidth}
    	    \centering
    		\includegraphics[clip, trim=2cm 1cm 2.5cm 1cm, width=0.89\textwidth]{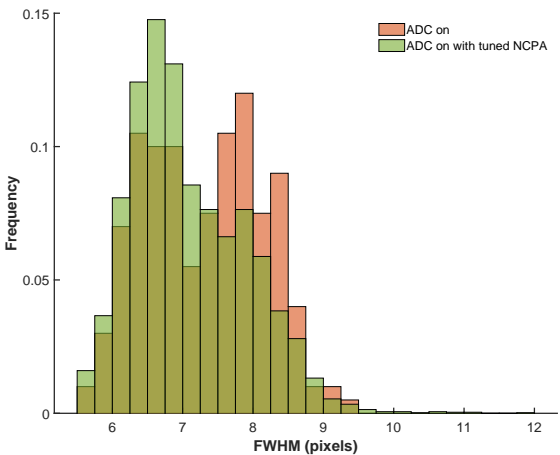}
		    \caption{Normalized histogram of PSF FWHM in the Y-axis of the ANDOR camera. NCPA correction has been tuned in one dataset (green) and standard values adopted in the other (orange). Both datasets have ADC correction applied.}
    \label{fig:adc_ncpa}
    \end{subfigure}
    \caption{Normalized histograms of PSF FWHM on the ANDOR camera for the target 2 Cygni which has been used to assess ADC performance.}
    \label{fig:adc_testing}
\end{figure}

High frame-rate data was recorded on the target 2 Cygni ($V_{\rm{mag}}=4.98$) at an airmass of 1.57 using the instrument ANDOR camera. Two sets of 200 frames were recorded, one with ADC correction enabled and one without. A histogram of measured FWHM in the X- and Y- direction on the detector for each dataset is shown in Fig.~\ref{fig:adc_testing}. With no ADC correction, the mean FWHM with 1$\sigma$ confidence interval is $6.91\substack{+0.65 \\ -0.43}$ pixels and $8.79\substack{+0.52 \\ -1.09}$ pixels in the X- and Y-direction respectively. The majority of dispersion is in the Y-direction as expected from the system design. 

When correction is enabled, the mean FWHM decreases to $6.51\substack{+0.98 \\ -0.45}$ pixels in X and $7.31\substack{+0.85 \\ -1.02}$ pixels in Y. With optimal ADC correction, the PSF should be circularly symmetric with equal FWHM values in X and Y. With correction enabled, the difference between the measured X- and Y-FWHM values is significantly reduced however, a small difference remains. This may indicate slight residual dispersion. Neither FWHM value reached the theoretical FWHM value however (6.1 pixels) which suggests the presence of NCPA and imperfect AO correction. This is consistent with the need for high airmass during these observations leading to reduced AO performance and standard NCPA values for higher elevations being used.

Tuning of NCPA for subsequent observations of the same target showed performance approaching the design limit (Fig.~\ref{fig:adc_ncpa}) with consistent mean FWHM values in the X and Y axes of $6.91\substack{+0.66 \\ -0.74}$ and $7.07\substack{+1.07 \\ -0.62}$ pixels respectively. This indicates the ADC is performing as designed even at low telescope elevations. Further assessment of the ADC performance using a wider imaging bandpass and its impact on fiber coupling efficiency is planned for future observing runs.

\subsection{Beam Stabilization Performance}
\label{sec:beam_stab_onsky}
Beam stabilization performance was characterized using a combination of ANDOR camera PSF centroid motion and fiber coupling efficiency. This method allows a direct assessment of the end-to-end stabilization system and its overall improvement on PSF positional stability. Data was recorded on several targets to directly assess performance of the control loop. PID parameters were tuned for each target to provide optimal beam stability. On certain targets, datasets were recorded with and without correction to allow for direct comparison of performance. 

\begin{figure}
    \centering
    \includegraphics[clip, trim=2.5cm 1cm 2.5cm 1cm, width=1.0\columnwidth]{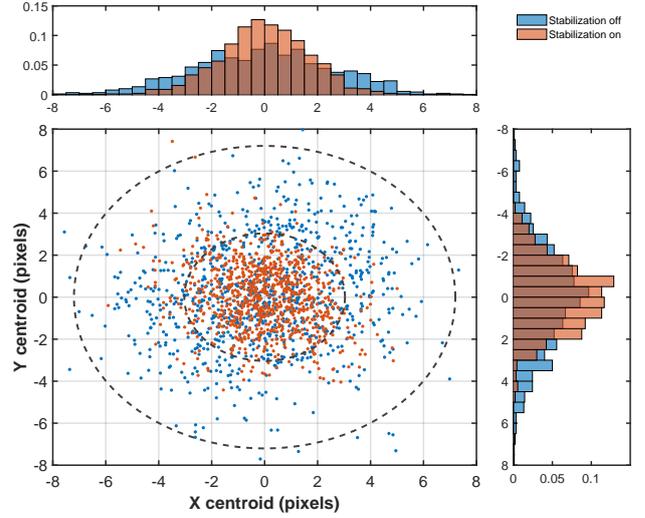}
    \caption{Location of the PSF centroid relative to the median location in the imaging channel for the target 2 Cygni with and without beam-stabilization enabled. 900 frames were recorded in each dataset. The nominal PSF scale in the imaging channel is represented by the dashed circles with the outer representing the first minimum and the inner representing the FWHM. Horizontal and vertical histograms are plotted to show the overall distribution. With stabilization enabled, the RMS deviation is reduced from 2.58 to 1.66 pixels in X and 2.48 to 1.59 pixels in Y.}
    \label{fig:quad_imaging_testing}
\end{figure}

Imaging data was recorded for the target 2 Cygni ($J_{\rm{mag}}=5.15$) with and without beam stabilization enabled  (Fig.~\ref{fig:quad_imaging_testing}). The mean distance of the centroid from the median location in all frames was calculated to be $3.18\substack{+1.80 \\ -1.60}$ without correction and $2.02\substack{+1.18 \\ -0.92}$ with correction enabled. These values are consistent with the performance noted during laboratory testing. With optimized correction, the stability of the PSF position was significantly improved and large-scale motion beyond 1 FWHM distance almost entirely eliminated.

\begin{figure}
    \centering
    \includegraphics[clip, trim=2.0cm 1cm 2.5cm 1cm, width=1.0\columnwidth]{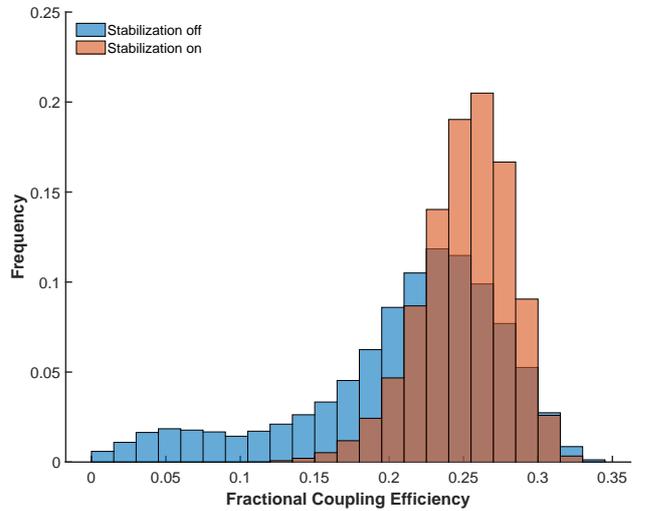}
    \caption{Normalized Histogram of the SMF coupling efficiency for target HD200253 with and without beam-stabilization enabled.}
    \label{fig:quad_coupling_testing}
\end{figure}

Similar performance was noted in fiber coupling data with \SI{140}{\sec} of data being recorded on the target HD~200253 ($J_{\rm{mag}}=4.68$) (Fig.~\ref{fig:quad_coupling_testing}). For the unstabilized case, the mean fractional coupling efficiency was measured to be $0.21\substack{+0.05 \\ -0.08}$ with an extended lower tail on the distribution correlating with the larger centroid motions noted in the previous imaging data. With beam stabilization enabled, the distribution of coupling efficiency narrows and the mean value increases to $0.25\substack{+0.03 \\ -0.03}$

The quad-cell detector and electronics limits were assessed by observing a series of targets with gradually increasing J-band magnitude to match the bandpass of the beam-stabilization channel. As expected from laboratory testing, PID parameters needed to be re-optimized with increasing magnitude and changing AO system performance. By monitoring the voltage of the quad-cell signal in parallel with assessing when no improvement in PSF stability was noted on the ANDOR camera when stabilization was enabled, the limiting magnitude of the sensor was found to be $J_{\rm{mag}}\approx7.5$. 

\subsection{Fiber Coupling Performance}

Fiber coupling performance has been assessed for a range of magnitudes and spectral types. For each target, the instrument MMF was positioned into the beam at the fiber focal plane and throughput measured with the FWR to provide a baseline of incident power. The SMF was then positioned into the beam and its precise location optimized using an iterative scanning approach until maximum coupling was achieved (\cref{sec:fiber_coupling}). The time taken to complete the coupling process was typically 5-10 minutes per target and was dependent upon the necessity of back-illumination for coarse fiber alignment.

Coupling of the full instrument bandpass was assessed in addition to measurements of coupling for narrower windows within the band to identify any chromatic performance effects. These narrowband measurements were obtained by inserting appropriate filters into the beam within the FWR. The telescope pupil geometry limits the maximum coupling to 78\% and simulations of on-sky performance estimated coupling to be $\sim$20\% in median seeing conditions.

\begin{figure}
    \centering
    \includegraphics[clip, trim=2.0cm 1cm 1.0cm 1cm, width=1.0\columnwidth]{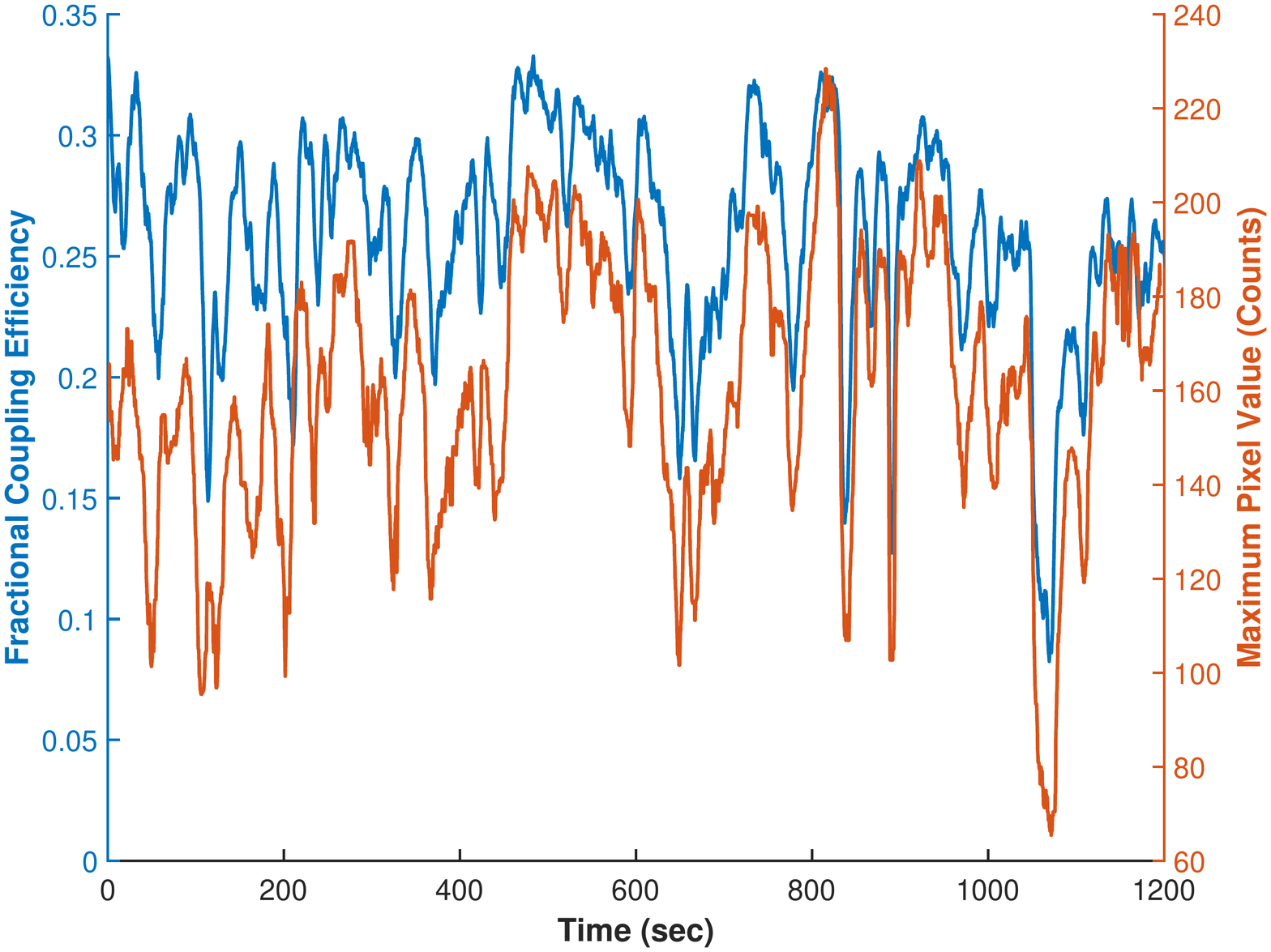}
    \caption{Simultaneous broadband (0.97-\SI{1.31}{\micro\meter}) fiber coupling efficiency (blue) plotted with intensity (counts) on the ANDOR camera (orange) for a 20 minute observation of HIP 28634. The maximum pixel value is used as a measure of intensity and as a proxy for AO system performance. The datasets were resampled to the same timebase and cross-correlated to calculate the offset between them. This offset has been applied to synchronize the two datasets in the plot with a median moving average filter being applied to remove high-frequency fluctuations in the figure. The mean fractional coupling achieved was 0.25$\pm$0.07 with a maximum value of 0.44.}
    \label{fig:fiber_pixel_coupling}
\end{figure}

The fractional coupling efficiency is strongly correlated with AO system performance (Fig.~\ref{fig:fiber_pixel_coupling}) as expected from theory and noted in the prototype system \citep{Bechter2020}. AO performance is also noted to correlate with recorded telescope seeing for many datasets.

\subsubsection{Broadband Performance}

\begin{figure}
    \centering
    \includegraphics[clip, trim=2.0cm 1cm 1.5cm 1cm, width=1.0\columnwidth]{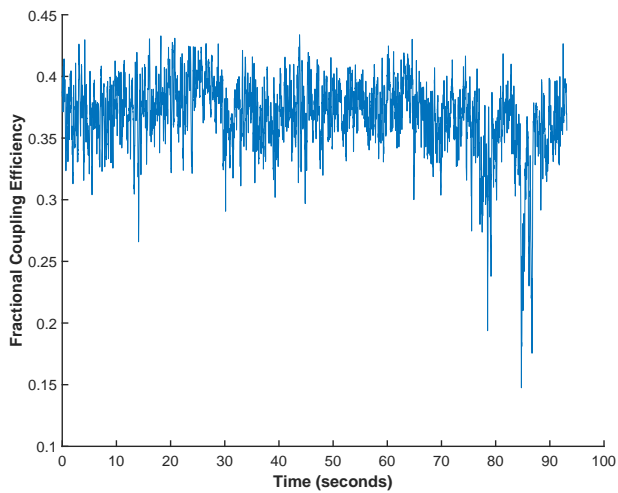}
    \caption{Fiber coupling efficiency for a 90 second observation of HIP~28634 (F5, $V_{\rm{mag}} = 6.49$). A moving average filter has been applied to remove the effects of high-frequency fluctuations.}
    \label{fig:HIP28634_max_coupling}
\end{figure}

\begin{figure}
    \centering
    \includegraphics[clip, trim=2.0cm 1cm 1.5cm 1cm, width=1.0\columnwidth]{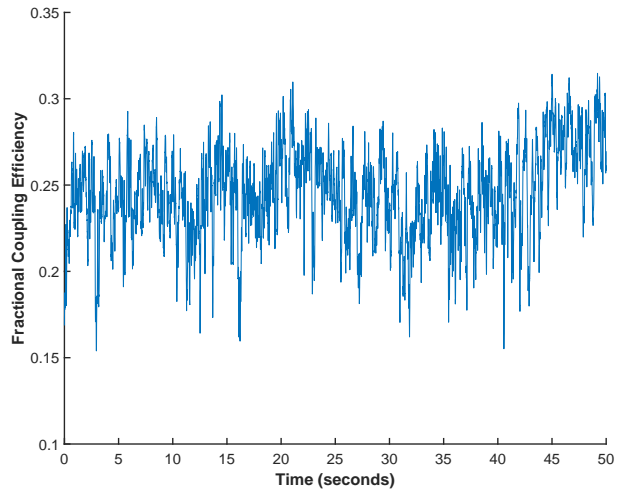}
    \caption{Fiber coupling efficiency for a 50 second observation of UCAC2 30573218 (M4II, $R_{\rm{mag}} = 13.50$). A moving average filter has been applied to remove the effects of high-frequency fluctuations.}
    \label{fig:UCA_coupling}
\end{figure}

Broadband fractional coupling efficiency for the F5 star HIP~28634 ($V_{\rm{mag}} = 6.49$) is shown in Fig.~\ref{fig:HIP28634_max_coupling}. NCPA correction had been optimized prior to this observation and the mean fractional coupling efficiency achieved was 0.37$\pm$0.03. The maximum fractional coupling recorded was 0.46. A degradation in performance is noted after \SI{70}{\sec} which corresponds to instability in AO correction. Broadband coupling was also recorded for the M4 star UCAC2 30573218 ($R_{\rm{mag}} = 13.50$) with a mean coupling efficiency of 0.24$\pm$0.04 and maximum coupling of 0.35 being achieved (Fig.~\ref{fig:UCA_coupling}). For all targets, coupling was maintained without manual adjustment of the acquisition camera system even during times of fluctuating AO system performance. Closed-loop position holding was enabled for the fiber stage. Datasets with sustained coupling of up to 20 minutes were recorded (Fig.~\ref{fig:fiber_pixel_coupling}).

Additional targets were observed with comparable coupling efficiencies being achieved. High frequency fluctuations were noted in all data sets with a significant difference between the average coupling value and the maximum value recorded ($\sim$10\% absolute). This is likely due to high-frequency tip/tilt residuals still being present in the beam reaching the fiber focal plane and causing the PSF to deviate from the optimal position on the fiber. These observations are consistent with estimated Strehl ratios recorded on the ANDOR camera which were in excess of 50\% for a range of targets. The Strehl ratio at the fiber focal plane is expected to exceed this value as it operates at a longer wavelength. Combined, this suggests a maximum coupling value $\sim$50\% which is consistent with the maximum value recorded, however, the system performance is reduced due to the beam constantly moving over the fiber tip.

An additional limit on performance was reached when the beam-stabilization loop was unable to close on the target due to its magnitude exceeding in the quad-cell sensitivity (\cref{sec:beam_stab_onsky}). Due to the residual beam motion in this situation, this precluded sufficient flux reaching the FWR for full optimization of fiber positioning. Observations exceeding the magnitude limited occurred primarily for earlier type stars where peak spectral emission is at visible wavelengths. The system design is optimized for late-type stars performance where peak emission is in the NIR and these targets were successfully coupled even at fainter AO guide-star magnitudes (Fig.~\ref{fig:UCA_coupling}).

\subsubsection{Narrowband Performance}

\begin{figure}
    \centering
    \includegraphics[clip, trim=2.0cm 1cm 1.5cm 1cm, width=1.0\columnwidth]{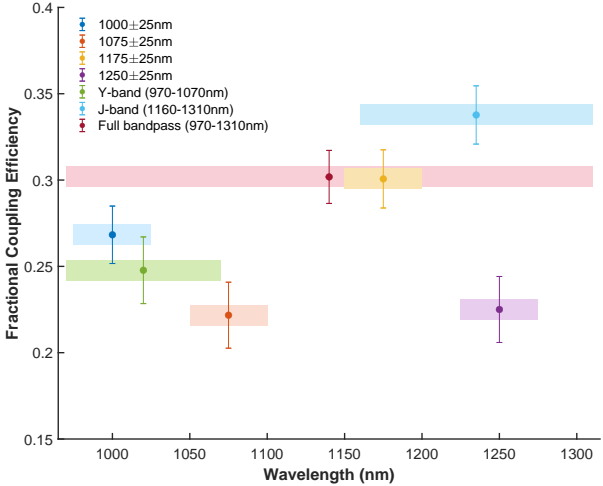}
    \caption{Narrowband fiber coupling efficiency over the iLocater bandpass for target HR~5553. Each datapoint represents the mean coupling efficiency. The width of each horizontal bar represents the bandwidth of the filter used for the measurement.}
    \label{fig:HR5553_bbandcoupling}
\end{figure}

In addition to broadband coupling of targets, narrowband measurements across the iLocater bandpass were recorded for the K0.5V star HR~5553 ($R_{\rm{mag}} = 5.50$) (Fig.~\ref{fig:HR5553_bbandcoupling}). The system was optimized for the first observation in the series and positions then held with no manual adjustment being made for subsequent data in the series. The entire dataset spans 18 minutes and seeing conditions were variable during this time (1.3$\pm$\ang{;;0.22}) leading to varying AO performance. Observations were paused for a period of 9 minutes during the set due to seeing increasing to a peak of \ang{;;1.85} before recovering to the previous levels. Despite the variable conditions, broadband coupling in the band exceeded 20\% for all wavelengths. A slight correlation in coupling efficiency is noted with wavelength however this is likely an effect of AO performance at the precise time each dataset was taken which is supported by the difference in coupling for observations with overlapping wavelength coverage (J-band, \SI{1075}{\nano\meter} and \SI{1250}{\nano\meter}). Further narrowband coupling characterization is planned for future observing runs. 

\subsection{Daytime Testing}

Daytime observations recorded since on-sky commissioning have been used to assess longer-term performance of the acquisition camera at the LBT. These observations have utilized illumination of the system from the calibration channel to measure internal performance. Testing was undertaken without beam-stabilization as the illumination source falls outside the beam-stabilization bandpass. A slow drift was noted in fiber coupling efficiency over time which was confirmed by assessing PSF centroid motion in the ANDOR channel. 

Given the elements in the optical path and the telescope remained at zenith during these tests, the suspected source of this perturbation is the thermal expansion of the tip/tilt mirror stage in the system. The stage is fabricated from stainless steel and any change in temperature will adjust the mirror position relative to the beam. While this effect is small, it is expected to be sufficient to induce motion in the fiber focal plane and decrease fiber coupling efficiency. Additional temperature monitoring hardware is being added to the instrument control system to confirm this correlation with temperature. For on-sky operation, any beam movement from this effect will be sensed by the quad-cell and corrected as part of the beam-stabilization control.

\section{Conclusions}
\label{sec:conclusions}

Mean on-sky SMF coupling efficiency exceeding 35\% (absolute) in the NIR (\SIrange{0.97}{1.31}{\micro\meter}) was demonstrated using the iLocater SX acquisition camera at the LBT. Fiber coupling was characterized in both broadband and narrowband observations with a maximum achieved SMF coupling of 46\% (absolute). This represents 59\% of the theoretical maximum coupling set by the telescope pupil geometry. Coupling efficiencies were improved by greater than a factor of two compared to the performance of the previously installed prototype system \citep{Bechter2020}.

The acquisition camera ADC and active beam-stabilization system performance were assessed using a combination of PSF profile, PSF centroid position, and fiber coupling efficiency. The ADC matched expected design performance, mitigating dispersion effects and PSF elongation even at low telescope elevations (\ang{35;;}). Beam stabilization reduced scatter in PSF centroid motion by 35\% and maintained position within 1 FWHM to ensure sustained fiber coupling during observations. Performance is consistent with laboratory testing prior to on-sky observations. The limiting magnitude of the beam-stabilization system was assessed as $J_{\rm{mag}}\approx7.5$.

Additional telescope observations were planned in 2020 to continue the performance assessment and optimization of the system, however, these have been postponed. Future observations will include verification of ADC performance using a wider imaging bandpass combined with fiber coupling measurements. Using narrowband filters in the FWR, wavelength-dependent fiber coupling can be studied, which will allow the identification of any wavelength-dependent PSF elongation in the fiber focal plane. Additional work is planned to ensure that optimal NCPA correction is applied for each observation.

To increase the limiting magnitude of the beam-stabilization system, upgrades to the electronics gain and sensitivity of the quad-cell detector are planned. Work is also on-going to identify pathways to increase the correction frequency of the system, which will in turn improve the positional stability of the PSF and reduce high-frequency fluctuations seen in fiber coupling data. Additionally, the optical throughput of the system will be increased by replacing current COTS mirrors with multi-layer coated equivalents.

Overall performance of the SX acquisition camera meets and exceeds the science requirements for the iLocater spectrograph. Fiber coupling performance is well optimized for bright stars as well as late-type stars. We find that fiber coupling efficiency is ultimately limited by AO performance and tip/tilt residuals on the incident telescope beam. Through beam-stabilization upgrades, tuning of NCPA corrections and optimized observations for the seeing conditions, the fiber injection system is well-positioned to support delivery of the iLocater spectrograph.

\section*{Acknowledgements}

The authors would like to thank Robert J. Harris, Philipp Hottinger and Jochen Heidt from the Landessternwarte K\"onigstuhl (Heidelberg University) for their support of the November 2019 observing run. The authors would also like to thank Grant West for his support with technical aspects of LBTI. JRC acknowledges support from NASA's Early Career program (NNX13AB03G) and the NSF CAREER Fellowship (NSF 1654125). We are grateful for support from the Potenziani family and the Wolfe family. 

The LBT is an international collaboration among institutions in the United States, Italy and Germany. LBT Corporation partners are: The University of Arizona on behalf of the Arizona Board of Regents; Istituto Nazionale di Astrofisica, Italy; LBT Beteiligungsgesellschaft, Germany, representing the Max-Planck Society, The Leibniz Institute for Astrophysics Potsdam, and Heidelberg University; The Ohio State University, representing OSU, University of Notre Dame, University of Minnesota and University of Virginia. 

This research has made use of the Washington Double Star Catalog maintained at the U.S. Naval Observatory.


\section*{Data Availability}

The data underlying this article will be shared on reasonable request to the corresponding author.




\bibliographystyle{mnras}

\bibliography{refs}






\bsp	
\label{lastpage}
\end{document}